\RequirePackage{iftex}
\ifPDFTeX
  \documentclass[11pt,a4paper]{amsart}
  \usepackage[margin=27mm]{geometry}
  \usepackage[T1]{fontenc}
  \usepackage{lmodern}
\else
  \documentclass[11pt, a4paper]{article}
  \usepackage[a4paper, margin=2.5cm]{geometry}
  \usepackage{fontspec}
  \usepackage[english, bidi=basic, provide=*]{babel}
  \babelprovide[import, onchar=ids fonts]{english}
  \babelfont{rm}{Noto Sans}
\fi
\usepackage{amsmath,amssymb,amsthm,mathtools,mathrsfs}
\ifPDFTeX
  \usepackage{microtype}
\fi
\usepackage{booktabs,tabularx}
\usepackage[hidelinks]{hyperref}

\allowdisplaybreaks
\numberwithin{equation}{section}

\newtheorem{theorem}{Theorem}[section]
\newtheorem{lemma}[theorem]{Lemma}
\newtheorem{proposition}[theorem]{Proposition}
\newtheorem{corollary}[theorem]{Corollary}
\theoremstyle{remark}
\newtheorem{remark}[theorem]{Remark}

\newcommand{\R}{\mathbb R}
\newcommand{\one}{\mathbf 1}
\newcommand{\cE}{\mathcal E}
\newcommand{\cX}{\mathcal X}
\ifPDFTeX
  \newcommand{\dbar}{{\mathchar'26\mkern-12mu\mathrm{d}}}
\else
  \newcommand{\dbar}{\text{\normalfont\dj}}
\fi

\begin{document}
\ifPDFTeX
  \title[Two correlation scales in density-matrix functionals]{Two correlation scales in homogeneous density-matrix functionals: the M\"uller logarithm and the sharp Cs\'anyi--Arias asymptotics}
  \author{Heinz Siedentop}
  \address{Mathematisches Institut\\ Ludwig-Maximilians-Universit\"at M\"unchen\\ Theresienstr. 39\\ 80333 M\"unchen\\ Germany}
  \email{h.s@lmu.de}
  \thanks{The work has been supported by the Deutsche Forschungsgemeinschaft (DFG TRR 352), grant 470903074. The author acknowledges the use of OpenAI's ChatGPT and Google's Gemini for mathematical discussions, the development and checking of arguments, and assistance in preparing the manuscript. The author takes full responsibility for the final text and proofs.}
  \date{13 November 2026}
\else
  \title{Two correlation scales in homogeneous density-matrix functionals:\\
  the M\"uller logarithm and the sharp Cs\'anyi--Arias asymptotics}
  \author{Heinz Siedentop\thanks{The work has been supported by the Deutsche Forschungsgemeinschaft (DFG TRR 352), grant 470903074. The author acknowledges the use of OpenAI's ChatGPT and Google's Gemini for mathematical discussions, the development and checking of arguments, and assistance in preparing the manuscript. The author takes full responsibility for the final text and proofs.}\\[1ex]
  \small Mathematisches Institut, Ludwig-Maximilians-Universit\"at M\"unchen,\\[-0.3ex]
  \small Theresienstr.\ 39, 80333 M\"unchen, Germany\\[-0.3ex]
  \small \texttt{h.s@lmu.de}}
  \date{13 November 2026}
  \maketitle
\fi

\begin{abstract}
We resolve the exact high-density correlation scales and coefficients of
the three-dimensional translation-invariant, spin-symmetric M\"uller and
Cs\'anyi--Arias corrected Hartree--Fock (CHF) functionals, settling
long-standing numerical observations of M\"uller overcorrelation and
apparently vanishing CHF correlation. In Hartree units, for fixed spin
multiplicity $q$ and Fermi momentum $k=(6\pi^2\rho/q)^{1/3}$, their gains
per particle over the filled Fermi sea satisfy
\[
 D_{\rm M}(k)=\frac{3}{4\pi^2}\log k+O(1),\qquad
 D_{\rm CHF}(k)=12e^{-1/2-\pi^2/8}
 k^2e^{-\pi^{3/2}\sqrt{k}}(1+o(1)).
\]
The M\"uller proof retains the occupation cap and identifies the
factor-of-two cancellation between exchange defect and energy gain.
For periodic M\"uller jellium, convexity and translation--spin averaging
allow a global minimizer to be chosen invariant and identify the
unrestricted thermodynamic energy with the homogeneous spin-symmetric
value. The CHF upper bound follows
from a deterministic comparison built from the explicit Hartree--Fock
dispersion, combining a full Birman--Schwinger bound with one-sided norms
and completion of the square; a number-preserving trial state provides
the matching lower bound, and the retained remainder yields weighted
$L^2$ convergence of the occupation defect without analyzing the
self-consistent gap. We also prove existence for both functionals and
uniqueness of the CHF minimizer. Relative to the unpolarized quantum
high-density benchmark, M\"uller has a larger logarithmic coefficient,
providing quantitative evidence for the conjectured M\"uller lower-bound
property, whereas CHF misses the logarithmic scale altogether. The quantum
benchmark is not used in either functional proof.
\end{abstract}
\ifPDFTeX
  \maketitle
\fi

\setcounter{tocdepth}{1}
\tableofcontents

\section{Introduction: origins, benchmarks, and the two correlation scales}
\label{sec:intro}
Reduced density-matrix functionals aim to retain the exact one-body
kinetic energy while replacing the two-body interaction by an explicit
expression in the one-particle density matrix. This places fractional
natural-orbital occupations, rather than an idempotent Hartree--Fock
projector, among the variational degrees of freedom. The foundational
background includes Coleman's characterization of ensemble-representable
one-particle density matrices~\cite{Coleman} and Gilbert's extension of
the density-functional framework to nonlocal external
potentials~\cite{Gilbert}. The conditions $0\le\gamma\le1$ and
$\operatorname{tr}\gamma=N$ concern one-particle representability; they
must not be confused with representability of an approximate
two-particle density matrix.

\subsection{From the square-root ansatz to corrected Hartree--Fock}
In 1984 M\"uller~\cite{Muller} proposed an approximation to the two-particle
density matrix that replaces the Hartree--Fock occupation factor
$n_i n_j$ in the exchange term by $\sqrt{n_i n_j}$. The square-root choice
is tied to the contraction relation between reduced density matrices
and to Pauli-type constraints. Buijse and Baerends~\cite{BB} later
obtained the same square-root expression from an exchange-correlation
hole construction. These are related routes to the functional now
usually called the M\"uller functional.

Cs\'anyi and Arias~\cite{CA} developed a tensor-product approximation to
the two-particle density matrix. Their ``corrected Hartree'' (CH)
functional coincides with the M\"uller functional. Their distinct
``corrected Hartree--Fock'' (CHF) functional retains ordinary exchange
and adds an exchange-type term involving $\sqrt{\gamma(1-\gamma)}$.
For homogeneous spin-symmetric occupations the two interaction factors
are, respectively,
\[
 \sqrt{n(p)n(p')},\qquad
 n(p)n(p')+
 \sqrt{n(p)(1-n(p))n(p')(1-n(p'))}.
\]
Throughout this paper, CHF denotes this original corrected
Hartree--Fock functional, not the later
Cs\'anyi--Goedecker--Arias functional~\cite{CGA}.
Staroverov and Scuseria~\cite{SS} related CHF optimization to
Hartree--Fock--Bogoliubov methods, with a reversal of the pairing-energy
sign. The associated gap equation is a useful variational structure;
it is not, by itself, a statement about superconductivity of the
repulsive many-electron Coulomb Hamiltonian.

\subsection{What was known about these functionals}
Cioslowski and Pernal~\cite{CP} analyzed homogeneous momentum
occupations and the regime in which the occupation cap is inactive.
Cs\'anyi and Arias~\cite{CA} numerically observed excess correlation for
CH and near-zero correlation for CHF at high density, reporting a
fitted logarithmic slope for CH in a study covering $0.01<r_s<30$.
Lathiotakis, Helbig, and Gross~\cite{LHG} subsequently compared these
and other density-matrix functionals on the homogeneous electron gas,
confirming both occupation pinning for M\"uller and the rapid collapse
of the CHF correlation energy as $r_s\downarrow0$, while
Cioslowski, Ziesche, and Pernal~\cite{CZP} derived high-density momentum
constraints on simple natural-spin-orbital functionals. The earlier
finite-density calculations reveal this qualitative contrast, but they
do not determine the exact high-density scales and coefficients with
controlled asymptotic errors. For M\"uller, a finite-range logarithmic
fit does not identify the limiting coefficient. For CHF, the
stretched-exponential gain poses a severe resolution problem,
particularly when it is obtained by subtracting total energies.
The calculations in~\cite{CA} discuss discretization, momentum-cutoff,
regularization, and minimization errors; \cite{LHG} also discusses
numerical false pinning of occupations. These issues are not limitations
of every possible numerical method: higher precision, suitable
rescaling, and a direct calculation of energy differences can improve
numerical access. The present analysis determines the
\emph{exact asymptotic scales and sharp limiting coefficients} without
relying on finite-range extrapolations.

The mathematical theory of M\"uller's functional was developed by
Frank, Lieb, Seiringer, and Siedentop~\cite{FLSS}: among other results,
they established convexity, uniqueness of the minimizing spatial
density, and existence in the atomic and molecular binding regime
$N\le Z$. These are results for the general density-matrix problem,
not a claim of uniqueness of every minimizing density matrix.
Siedentop~\cite{Siedentop2009} showed that the neutral-atom M\"uller
energy agrees with the quantum energy through the Thomas--Fermi,
Scott, and Dirac--Schwinger orders. More recently,
Siedentop~\cite{Siedentop} proved the general comparison
$\mathcal E_{\rm M}\le\mathcal E_{\rm CHF}\le\mathcal E_{\rm HF}$
and deduced the same atomic accuracy for CHF. Agreement through these
large atomic energy orders leaves smaller correlation terms unresolved.
It does not imply agreement of homogeneous correlation energies.

\subsection{Why the high-density comparison is discriminating}
In the homogeneous setting, the common filled-Fermi-sea energy per
particle is
\[
 e_{\rm HF}^{\rm hom}(k)=\frac{3}{10}k^2-\frac{3}{4\pi}k.
\]
Both approximations recover these kinetic and exchange terms. Our
results show, however, that their next energy scales are radically
different. The M\"uller gain is
$3(4\pi^2)^{-1}\log k+O(1)$, whereas the CHF gain is
\[
 12e^{-1/2-\pi^2/8}\,k^2e^{-\pi^{3/2}\sqrt{k}}(1+o(1)).
\]
In particular, the CHF correction is positive at every finite density
but smaller than every inverse power of $k$ at high density. This
resolves the scale hidden behind a numerical observation of
``essentially zero'' correlation. In the M\"uller case, identifying the
coefficient, rather than only logarithmic growth, tests the quantitative
quality of its correlation mechanism.

For physical, unpolarized electrons, the high-density quantum benchmark
is the Macke--Gell-Mann--Brueckner logarithm. Macke~\cite{Macke} and
Pines~\cite{Pines} obtained the leading logarithmic behavior; Gell-Mann
and Brueckner~\cite{GMB} developed the systematic resummation and the
next constant term. In Hartree units the logarithmic coefficient is
$(1-\log2)/\pi^2$; see also Loos and Gill~\cite{LG} for the normalization
and the role of the Hartree--Fock reference. Consequently M\"uller has
the right logarithmic scale but a different coefficient, while CHF has
a different scale altogether. Neither the functional ordering nor
agreement of leading total energies can detect this distinction.

A complementary motivation is the conjecture that the M\"uller
ground-state energy bounds the exact quantum ground-state energy from
below for Coulomb systems~\cite{FLSS,Siedentop}. From this perspective,
the larger logarithmic correlation coefficient is not merely an
inaccuracy of approximation: relative to the quantum benchmark it
places the M\"uller energy on the conjectured side, with a strictly
positive leading separation. Section~\ref{sec:mueller-lower-bound}
makes this comparison quantitative. It supplies evidence for the
conjecture, without proving it for general Coulomb systems.

The mathematical status of this comparison requires care. The results
below are theorems about two specified homogeneous variational
functionals. Christiansen, Hainzl, and Nam~\cite{CHNupper,CHNlower}
proved a Coulomb Gell-Mann--Brueckner analogue in a mean-field regime.
That result must not be identified with the unscaled thermodynamic
jellium asymptotic. We keep this distinction explicit in
Section~\ref{sec:quantum-comparison}; the latter formula is used as the
physical benchmark, not as an input to either functional proof.

\subsection{A direct variational proof}
The CHF upper bound is a comparison for every admissible occupation,
not a consequence of an asymptotic solution of the gap equation.
We choose an explicit comparison energy $E_*$ whose Coulomb
Birman--Schwinger operator has norm at most one. The norms restricted
to either side of the Fermi ball tend to one half; this additional
fact controls the ordinary exchange of the signed occupation defect.
Completion of the square then leaves a positive weighted $L^2$
remainder. Thus the energy estimate also controls profiles of approximate
minimizers. A trial state on the scale selected by the same deterministic
spectrum supplies the matching lower bound.

There is also an important structural difference regarding symmetry
restrictions. In Hartree--Fock theory, the energy improvement due to
translation-symmetry breaking is exponentially small at high
density~\cite{GHL}. Our CHF analysis remains in the homogeneous
spin-symmetric class: we neither remove this restriction nor prove
symmetry breaking for CHF. For the M\"uller functional, by contrast,
convexity with respect to the one-particle density matrix~\cite{FLSS}
implies that averaging over translations and $\mathrm{U}(q)$ spin
rotations cannot increase the periodic energy and eliminates the direct
Coulomb term. Proposition~\ref{prop:M-periodic} proves existence of an
invariant global periodic minimizer and identifies the unrestricted
thermodynamic energy with the homogeneous spin-symmetric value.
This energy reduction does not assert that every minimizing density
matrix is itself invariant.

After the common setting and Coulomb estimates,
Section~\ref{sec:mueller} proves the M\"uller theorem and its periodic
global optimality.
The CHF uniqueness result in Section~\ref{sec:CHF-structure} is
independent of its high-density energy proof.
Sections~\ref{sec:direct}--\ref{sec:energy-last} give the deterministic
comparison, its spectral input, and the matching trial calculation;
Section~\ref{sec:profile} deduces the defect profile.
Section~\ref{sec:quantum-comparison} discusses the quantum benchmark
and the lower-bound motivation. No pointwise high-density asymptotic
for the self-consistent CHF gap, or atomic local-density transfer, is
asserted.

\section{Common setting and main results}\label{sec:conventions}
We first formulate both variational problems in the translation-invariant,
spin-symmetric class. For M\"uller, Proposition~\ref{prop:M-periodic}
shows that this entails no loss in the periodic thermodynamic energy.
For CHF, the symmetry restriction remains part of the problem studied.
All units are Hartree units. Fix a spin multiplicity $q\in\mathbb N$ and a
positive total electron density $\rho$. We use the notation
$\dbar p=\mathrm{d}p/(2\pi)$ for the one-dimensional normalized measure,
and correspondingly
\[
 \dbar^3p=\frac{\mathrm{d}^3p}{(2\pi)^3},\qquad
 (Vf)(p)=\int_{\R^3}\frac{4\pi f(p')}{|p-p'|^2}\,\dbar^3p',\qquad
 \cX(f)=\frac q2\int f(Vf)\,\dbar^3p.
\]
All other differentials are written with an upright $\mathrm{d}$.
Unless indicated otherwise, $L^p$ norms are taken on $(\R^3,\dbar^3p)$,
whereas $\|\cdot\|$ on linear operators denotes the operator norm.
Fix once and for all $0<\theta<1/12$; the shell parameter
is independent of $k$, and no limit $\theta\downarrow0$ is needed.
Constants $C_\theta$ may depend on this fixed choice but not on $k$.
The Coulomb quadratic form $\cX$ is nonnegative even for real
sign-changing functions. Define
\begin{align}
 \cE_{\rm HF}(n)
 &=\frac q2\int_{\R^3}|p|^2n(p)\,\dbar^3p-\cX(n),
 \label{eq:HF-energy}\\*
 \cE_{\rm M}(n)
 &=\frac q2\int_{\R^3}|p|^2n(p)\,\dbar^3p-\cX(\sqrt n),
 \label{eq:M-energy}\\*
 \cE(n)=\cE_{\rm CHF}(n)
 &=\frac q2\int_{\R^3}|p|^2n(p)\,\dbar^3p
        -\cX(n)-\cX(w_n),\qquad w_n=\sqrt{n(1-n)}.
 \label{eq:energy}
\end{align}
The abbreviation $\cE$ without a subscript always denotes CHF. The
M\"uller energy in \eqref{eq:M-energy} is the unshifted functional;
adding a density-linear self-energy correction would only change its
per-particle remainder by a constant, not the logarithmic coefficient.
The admissible set is
\[
 \mathcal A_\rho=\left\{n:\R^3\to[0,1]:
 q\int n\,\dbar^3p=\rho,\quad \int|p|^2n\,\dbar^3p<\infty\right\}.
\]
For the general density-matrix comparison see~\cite{Siedentop}.
In the present scalar class the pointwise ordering can be checked directly:
for $s,t\in[0,1]$, Schwarz's inequality gives
\[
 st+\sqrt{s(1-s)t(1-t)}
 =\sqrt{st}\bigl(\sqrt{st}+\sqrt{(1-s)(1-t)}\bigr)
 \le\sqrt{st}.
\]
Therefore, on the same admissible set,
\begin{equation}\label{eq:functional-order}
 \cE_{\rm M}(n)\le\cE_{\rm CHF}(n)\le\cE_{\rm HF}(n).
\end{equation}
This is an ordering of approximate functionals, not an assertion about
where the exact many-body quantum energy lies.

The homogeneous direct Coulomb term is canceled by the neutralizing
background. This variational problem is \emph{not} unrestricted
Hartree--Fock; in particular we do not allow general spatially varying
one-body density matrices or spin-symmetry breaking in the CHF analysis.
For the M\"uller functional, however, convexity permits an invariant
global minimizer for periodic jellium, and restricting to invariant
states does not change the minimum
(Proposition~\ref{prop:M-periodic}).

Set
\begin{equation}\label{eq:fermi}
 k=\left(\frac{6\pi^2\rho}{q}\right)^{1/3},\qquad
 P_k(p)=\one_{\{|p|<k\}},\qquad
 D(k)=\frac{\cE(P_k)-\inf_{n\in\mathcal A_\rho}\cE(n)}\rho.
\end{equation}
For $\nu\in\{{\rm M},{\rm CHF}\}$ define the energy per particle
and the nonnegative gain by
\begin{equation}\label{eq:common-gains}
 e_\nu^{\rm hom}(k)=\rho^{-1}\inf_{n\in\mathcal A_\rho}\cE_\nu(n),
 \qquad D_\nu(k)=e_{\rm HF}^{\rm hom}(k)-e_\nu^{\rm hom}(k).
\end{equation}
At fixed $k$, the constraint
$\int n\,\dbar^3p=k^3/(6\pi^2)$ is independent of $q$.
Both the energy functionals and
$\rho=qk^3/(6\pi^2)$ carry the same factor $q$.
Consequently, the entire energies per particle and their gains are
independent of $q$ when expressed as functions of $k$:
\[
 e_\nu^{\rm hom}(k;q)=e_\nu^{\rm hom}(k;1),
 \qquad
 D_\nu(k;q)=D_\nu(k;1),
 \qquad \nu\in\{{\rm M},{\rm CHF}\}.
\]
The same observation applies to the Hartree--Fock reference energy;
it does not concern a change of $q$ at fixed total density.
The reference energy is common to all three functionals:
\begin{equation}\label{eq:common-HF}
 \rho^{-1}\cE_{\rm M}(P_k)
 =\rho^{-1}\cE_{\rm CHF}(P_k)
 =\rho^{-1}\cE_{\rm HF}(P_k)
 =e_{\rm HF}^{\rm hom}(k)
 =\frac{3k^2}{10}-\frac{3k}{4\pi}.
\end{equation}
For completeness, we spell out the exchange comparison. Let $n^\ast$
be the symmetric decreasing rearrangement of $n\in\mathcal A_\rho$ and set
\[
 \Phi=V(n^\ast+P_k).
\]
The function $\Phi$ is bounded and radially nonincreasing, since
$n^\ast+P_k\in L^1\cap L^\infty$ and both $n^\ast+P_k$ and the
Coulomb kernel are nonnegative and radially nonincreasing. Riesz
rearrangement, symmetry of the kernel, and
$\int(n^\ast-P_k)\,\dbar^3p=0$ give
\begin{equation}\label{eq:exchange-maximality}
\begin{aligned}
 \cX(n)-\cX(P_k)
 &\le\cX(n^\ast)-\cX(P_k)\\
 &=\frac q2\int_{\R^3}
   \bigl(n^\ast(p)-P_k(p)\bigr)
   \bigl[\Phi(|p|)-\Phi(k)\bigr]\,\dbar^3p\\
 &\le0.
\end{aligned}
\end{equation}
Indeed, the two factors in the integrand have opposite signs both
inside and outside the Fermi ball. All integrals are absolutely
convergent because $\Phi$ is bounded and $n^\ast-P_k\in L^1$.
Together with the bathtub principle, \eqref{eq:exchange-maximality}
shows that $P_k$ minimizes $\cE_{\rm HF}$ over the whole admissible
class $\mathcal A_\rho$.
In the CHF sections we retain the shorter notation
$D(k)=D_{\rm CHF}(k)$. By \eqref{eq:functional-order},
$0\le D(k)\le D_{\rm M}(k)$.

\begin{theorem}[The M\"uller logarithm]\label{thm:mueller-main}
For every fixed spin multiplicity $q\in\mathbb N$, the M\"uller
energy gain in \eqref{eq:common-gains} satisfies
\begin{equation}\label{eq:M-main}
 \boxed{D_{\rm M}(k)=\frac{3}{4\pi^2}\log k+O(1)
 \qquad(k\to\infty).}
\end{equation}
The bound on the remainder is independent of $q$. The statement
controls a bounded remainder but does not assert that this remainder
has a limit.
\end{theorem}

For the CHF asymptotics write
\begin{equation}\label{eq:CHF-constants}
 A=\frac{\pi^{3/2}}2,\qquad
 y_0=4e^{-1/4-\pi^2/16},\qquad
 C_*=\frac34y_0^2=12e^{-1/2-\pi^2/8}.
\end{equation}
\begin{theorem}[Sharp CHF high-density asymptotics]\label{thm:main}
For every fixed spin multiplicity $q\in\mathbb N$, the energy gain
$D(k)$ defined in \eqref{eq:fermi} satisfies
\begin{equation}\label{eq:constant-target}
 \boxed{\lim_{k\to\infty}
 \frac{D(k)}{k^2e^{-\pi^{3/2}\sqrt{k}}}
 =12\exp\left(-\frac12-\frac{\pi^2}{8}\right).}
\end{equation}
The reference energy is that of the filled Fermi sea in the
translation-invariant, spin-symmetric class. The limiting constant is
independent of $q$ and equals $2.1195548699\ldots$.
\end{theorem}

The comparison argument also gives the weighted defect-profile
convergence in Corollary~\ref{cor:L2-profile}, including an assertion
for approximate minimizers. Its scale is deterministic and is not
identified with the self-consistent gap.

\section{Coulomb estimates, existence, and radial coordinates}
\label{sec:common-tools}
\subsection{Coulomb estimates}
We shall use
\begin{align}
 0\le\cX(f)&\le Cq\|f\|_2\,\||p|f\|_2,
 \label{eq:dual-hardy}\\
 \|Vf\|_\infty&\le C\|f\|_\infty^{1/3}\|f\|_2^{2/3},
 \label{eq:M-field-sup}\\
 \int_{\R^3}\frac{|Vf(p)|^2}{|p|^2}\,\dbar^3p
 &\le4\|f\|_2^2.
 \label{eq:M-inverse-moment}
\end{align}
The first estimate follows from Schwarz and Hardy after Fourier
transformation, where $V$ is multiplication by $|x|^{-1}$.
For the second, split the convolution at radius $s>0$:
$\|Vf\|_\infty\le C(s\|f\|_\infty+s^{-1/2}\|f\|_2)$,
and optimize $s$. The same bound applies to signed functions by
using $|f|$. For the third,
$\|\nabla_pVf\|_2=\|f\|_2$, and Hardy in the $p$ variable applies.
This is a homogeneous Sobolev assertion; it does not require $Vf\in L^2$.
Approximation justifies these estimates in the stated norm classes.
The convolution of a nonnegative $L^2\cap L^\infty$ function is
bounded and continuous, and strictly positive if the function is
not identically zero.

\begin{lemma}[Continuity of the Coulomb form]\label{lem:compact}
Suppose $f_j\rightharpoonup f$ weakly in $L^2(\R^3,\dbar^3p)$ and
$\sup_j(\|f_j\|_2+\||p|f_j\|_2)<\infty$. Then
$\cX(f_j)\to\cX(f)$.
\end{lemma}
\begin{proof}
For $f_j^{\rm out}=\one_{\{|p|>s\}}f_j$,
$\|f_j^{\rm out}\|_2=O(s^{-1})$ and
$\||p|f_j^{\rm out}\|_2=O(1)$. Thus
\eqref{eq:dual-hardy} bounds its Coulomb form by $C/s$.
Schwarz for the positive Coulomb form controls the mixed term uniformly.
On a fixed ball the Coulomb operator is compact: cutting out
$|p-p'|<\delta$ leaves a Hilbert--Schmidt kernel, and Schur's test
bounds the removed operator by $C\delta$. Weak convergence proves
convergence of the interior quadratic forms. Now let $s\to\infty$.
The weak limit satisfies the same weighted bounds.
\end{proof}

\subsection{Existence for both functionals}
\begin{theorem}[Existence]\label{thm:exist}
For every $q\in\mathbb N$ and $\rho>0$, both $\cE_{\rm M}$ and
$\cE_{\rm CHF}$ attain their infima on $\mathcal A_\rho$.
\end{theorem}
\begin{proof}
Put $N=\rho/q$ and $T(n)=\int|p|^2n\,\dbar^3p$.
For either functional, $n^2\le n$ and $w_n^2\le n$ give
\begin{equation}\label{eq:coercive}
 \cE_\nu(n)\ge\frac q2T(n)-Cq\sqrt{NT(n)},
 \qquad \nu\in\{{\rm M},{\rm CHF}\}.
\end{equation}
For a minimizing sequence $n_j$ the kinetic moments are bounded;
in particular,
\begin{equation}\label{eq:tight}
 \int_{|p|>s}n_j\,\dbar^3p\le C/s^2.
\end{equation}
Set $z_j=\sqrt{n_j}$ for M\"uller and
$z_j=\sqrt{n_j(1-n_j)}$ for CHF. Extract weak $L^2$ limits
$n_j\rightharpoonup n$, $z_j\rightharpoonup z$.
The hypographs of the two concave functions
$\varphi(s)=\sqrt s$ and $\varphi(s)=\sqrt{s(1-s)}$ on $[0,1]$
are closed and convex. Weak closedness therefore implies
\begin{equation}\label{eq:relax}
 0\le n\le1,\qquad 0\le z\le\varphi(n)\quad\text{a.e.}
\end{equation}
Local weak convergence and \eqref{eq:tight} preserve $\int n=N$,
and $T(n)\le\liminf T(n_j)$. Lemma~\ref{lem:compact} applies to
$z_j$ and, for CHF, to $n_j$; all their squared weighted moments
are bounded by $T(n_j)$. Replacing $z$ by $\varphi(n)$ increases
its positive-kernel Coulomb form. This proves lower semicontinuity
in the required direction for either functional and hence existence.
\end{proof}

\subsection{Explicit radial formulas}
For radial real functions,
\begin{equation}\label{eq:radial-X}
 \cX(f)=\frac q{4\pi^3}\int_0^\infty\!\int_0^\infty
 rs f(r)f(s)\log\frac{r+s}{|r-s|}\,\mathrm{d}r\,\mathrm{d}s.
\end{equation}
The potential of the Fermi ball is
\begin{equation}\label{eq:fermi-field}
 (VP_k)(r)=\frac1\pi\left[
 k+\frac{k^2-r^2}{2r}\log\frac{k+r}{|k-r|}\right],
 \qquad (VP_k)(k)=\frac k\pi.
\end{equation}
The values at $r=0,k$ are taken by continuity. To preserve radial
particle number exactly, use the volume coordinate
\begin{equation}\label{eq:volume}
 t=\frac{r^3-k^3}{3k^2},\quad
 r(t)=(k^3+3k^2t)^{1/3},\quad
 \dbar^3p\ \text{(radially)}=\frac{k^2}{2\pi^2}\,\mathrm{d}t,
 \qquad t\ge-k/3.
\end{equation}
We identify a radial momentum-space function with its radial profile;
a tilde denotes its expression in the volume coordinate. In particular,
$h_k(p)=n_k(p)-P_k(p)$ and
$\widetilde h_k(t)=h_k(r(t))$. For nonradial competitors $h_k$ always
remains a function of $p\in\R^3$.

\begin{lemma}[Fixed-shell Coulomb bounds]\label{lem:shell-kernel}
Put $R=\theta k$. For $|t|,|s|\le R$, the kernel
\[
 H_k(t,s)=\frac{k^2}{r(t)r(s)}
           \log\frac{r(t)+r(s)}{|r(t)-r(s)|}
\]
satisfies
\begin{align}
 0\le H_k(t,s)&\le C_\theta
 \left(1+\log_+\frac{C_\theta k}{|t-s|}\right),
 \label{eq:shell-log-bound}\\
 \left|H_k(t,s)-\log\frac{2k}{|t-s|}\right|
 &\le C_\theta\frac{|t|+|s|}{k}
       \left(1+\log\frac{2k}{|t-s|}\right).
 \label{eq:shell-geometry}
\end{align}
If $0\le f\le1$ is supported in $[-R,R]$ and
$m=\int f\,\mathrm{d}t>0$, then
\begin{equation}\label{eq:log-mass-bound}
 \sup_t\int f(s)\left(1+\log_+\frac{C_\theta k}{|t-s|}\right)
 \,\mathrm{d}s
 \le C_\theta m\left(1+\log\frac{C_\theta k}{m}\right).
\end{equation}
\end{lemma}
\begin{proof}
Use $r(t)^3-r(s)^3=3k^2(t-s)$ and
$r(t)/k=1+O_\theta(|t|/k)$. In particular,
\[
 \frac{r(t)+r(s)}{|r(t)-r(s)|}
 =\frac{(r(t)+r(s))(r(t)^2+r(t)r(s)+r(s)^2)}
        {3k^2|t-s|}
 =\frac{2k}{|t-s|}
  \left(1+O_\theta\left(\frac{|t|+|s|}{k}\right)\right).
\]
This proves the two kernel estimates. For \eqref{eq:log-mass-bound}
split at $|t-s|=m$: use $f\le1$ inside and total mass $m$ outside.
The local logarithm is integrable, uniformly in the observation point.
\end{proof}

\section{The M\"uller logarithm with the occupation cap}
\label{sec:mueller}
This section proves Theorem~\ref{thm:mueller-main} in the common
normalization of Section~\ref{sec:conventions}, and then establishes
global optimality in periodic M\"uller jellium. Write
$b=\sqrt n$ and
\[
 \mathcal F_{\rm M}(b):=\cE_{\rm M}(b^2)
 =\frac q2\int |p|^2b(p)^2\,\dbar^3p-\cX(b),
 \qquad 0\le b\le1,\quad q\|b\|_2^2=\rho.
\]
The field is $\sigma_b=Vb$. The letter $b$ in this section denotes the
square root of the occupation.

\subsection{Existence and the constrained Euler equation}
\begin{lemma}\label{lem:M-variational}
The M\"uller variational problem has a minimizer. One may choose
$b_{{\rm M},k}$ radial and nonincreasing, and $b_{{\rm M},k}>0$ almost everywhere.
There is a real multiplier $\mu_{\rm M}$ such that, with
$g(p)=|p|^2/2-\mu_{\rm M}$,
\begin{equation}\label{eq:M-KKT}
 2g b_{{\rm M},k}=\sigma_{b_{{\rm M},k}}\quad\text{on }\{0<b_{{\rm M},k}<1\},
 \qquad 2g\le\sigma_{b_{{\rm M},k}}\quad\text{on }\{b_{{\rm M},k}=1\}.
\end{equation}
In particular $b_{{\rm M},k}=1$ where $g\le0$; where $g>0$,
\begin{equation}\label{eq:M-cap}
 b_{{\rm M},k}=\min\left\{1,\frac{\sigma_{b_{{\rm M},k}}}{2g}\right\}.
\end{equation}
\end{lemma}
\begin{proof}
Existence follows from Theorem~\ref{thm:exist}.
Symmetric decreasing rearrangement of $b$ preserves
its cap and norm, decreases the kinetic term and increases $\cX(b)$.
A radial nonincreasing minimizer can consequently be selected.

The potential $\sigma_b$ is positive and bounded for every nonzero
admissible $b$. If $n=b^2$ vanished on a set of positive measure, add
$\varepsilon$ to $n$ on a bounded subset and remove the same mass
from a bounded set on which $n$ is bounded below. The latter change
has energy cost $O(\varepsilon)$, whereas the new square-root
occupation produces a strictly negative energy change of order
$\sqrt\varepsilon$. This contradicts minimality. The same argument
rules out an indicator minimizer. Local mass-preserving variations
on $\{0<n<1\}$ give a common multiplier; one-sided variations on
$\{n=1\}$ give the indicated inequality. Exhaust the fractional set by
$\{|p|\le j,\ j^{-1}\le n\le1-j^{-1}\}$; overlapping
positive-measure subsets give the same multiplier. This proves
\eqref{eq:M-KKT} and \eqref{eq:M-cap}.
\end{proof}

\subsection{The pinned core and localization}
In the rest of this section $b=b_{{\rm M},k}$ denotes the selected minimizer,
and constants are independent of large $k$. Its cap and mass give
\begin{equation}\label{eq:M-field-Ck}
 0<\sigma_b(p)\le Ck.
\end{equation}
For sufficiently large $k$, we claim that
\begin{equation}\label{eq:M-mu}
 0\le\delta_k^{\rm M}:=\frac{k^2}{2}-\mu_{\rm M}\le Ck.
\end{equation}
If $\mu_{\rm M}\le0$, \eqref{eq:M-cap} would give
$b(p)^2\le\min\{1,Ck^2/|p|^4\}$ and therefore
$\|b\|_2^2\le Ck^{3/2}$, contradicting
$\|b\|_2^2=k^3/(6\pi^2)$ for large $k$. Thus
$\kappa=\sqrt{2\mu_{\rm M}}>0$. Since $b=1$ on $|p|<\kappa$,
mass conservation gives $\kappa\le k$. Outside this ball,
\[
 b(r)^2\le\min\{1,Ck^2/(r^2-\kappa^2)^2\}.
\]
With $v=r^2-\kappa^2$ and
$r^2\,\mathrm{d}r=\tfrac12\sqrt{\kappa^2+v}\,\mathrm{d}v$,
splitting at $v=k$ proves
\[
 \int_{|p|>\kappa}b(p)^2\,\dbar^3p
 \le C\int_0^\infty(\kappa+\sqrt v)
             \min\{1,k^2/v^2\}\,\mathrm{d}v
 \le C(k\kappa+k^{3/2})\le Ck^2.
\]
Thus $k^3-\kappa^3=O(k^2)$, which implies
$k-\kappa=O(1)$ and proves \eqref{eq:M-mu}.

Let $P=P_k$ and $H=b-P$. Equal masses imply
\begin{equation}\label{eq:M-localization}
 \int|b^2-P|\,\dbar^3p=O(k^2),\qquad
 \|H\|_2^2\le\int|b^2-P|\,\dbar^3p=O(k^2).
\end{equation}
Consequently \eqref{eq:M-field-sup} gives
\begin{equation}\label{eq:M-field-approx}
 \|\sigma_b-\sigma_P\|_\infty=O(k^{2/3}).
\end{equation}
This is an absolute error, uniform in momentum. It is sufficient
because its contribution to the final energy is $O(k^3)$.
Moreover, on $r>2k$, \eqref{eq:M-cap} and $2\mu_{\rm M}\le k^2$
give $r^2b(r)^2\le C\sigma_b(r)^2/r^2$. Using
\eqref{eq:M-inverse-moment} there and \eqref{eq:M-localization}
inside $r\le2k$, we obtain
\begin{equation}\label{eq:M-H-moments}
 \||p|H\|_2^2=O(k^4),\qquad
 0\le\cX(H)\le Cq\|H\|_2\||p|H\|_2=O(qk^3).
\end{equation}

\subsection{The logarithmic integral}
Set
\[
 Q_k=q\int_{|p|>k}g(p)b(p)^2\,\dbar^3p.
\]
We prove
\begin{equation}\label{eq:M-Q}
 Q_k=\frac{qk^3}{8\pi^4}\log k+O(qk^3).
\end{equation}
Choose a fixed $s_0>0$, independent of $k$, large enough that
$2g>\sigma_b$ on $r\ge k+s_0$; this is possible by
\eqref{eq:M-field-Ck}. On this region $b=\sigma_b/(2g)$.
The shell $k<r<k+s_0$ contributes $O(qk^3)$ by
\eqref{eq:M-mu}. The tail $r>2k$ also contributes $O(qk^3)$,
since $g b^2=\sigma_b^2/(4g)\le C\sigma_b^2/r^2$ and
\eqref{eq:M-inverse-moment} applies. Hence
\[
 Q_k=\frac{q}{8\pi^2}\int_{k+s_0}^{2k}
               \frac{r^2\sigma_b(r)^2}{g(r)}\,\mathrm{d}r
               +O(qk^3).
\]
Because $\sigma_b,\sigma_P=O(k)$, \eqref{eq:M-field-approx}
and $\int_{k+s_0}^{2k}r^2/g(r)\,\mathrm{d}r=O(k\log k)$
show that replacing $\sigma_b^2$ by $\sigma_P^2$ costs
$O(qk^{8/3}\log k)=O(qk^3)$.

Use the explicit Fermi-ball field \eqref{eq:fermi-field}.
Writing $r=k+s$, $s_0\le s\le k$, gives
\[
 |\sigma_P(k+s)-k/\pi|
 \le Cs\bigl(1+\log(k/s)\bigr),\qquad g(k+s)\ge ks.
\]
It follows that replacing $\sigma_P(r)^2$ by $k^2/\pi^2$
in the preceding integral costs $O(qk^3)$, since
$\int_{s_0}^k(1+\log(k/s))\,\mathrm{d}s=O(k)$.
Finally, putting $g_0=(r^2-k^2)/2$ and using
$g=g_0+\delta_k^{\rm M}$,
\[
 \left|\int_{k+s_0}^{2k}r^2
 \left(\frac1g-\frac1{g_0}\right)\,\mathrm{d}r\right|
 \le Ck\int_{s_0}^k s^{-2}\,\mathrm{d}s=O(k).
\]
The remaining elementary integral is
\[
 \int_{k+s_0}^{2k}\frac{r^2}{g_0(r)}\,\mathrm{d}r
 =2\int_{k+s_0}^{2k}\frac{r^2}{r^2-k^2}\,\mathrm{d}r
 =k\log k+O(k).
\]
This proves \eqref{eq:M-Q}, with all replacements controlled
at order $qk^3$.

\subsection{Cancellation in the energy and proof of the theorem}
The boundary multiplier must be retained. Define the residual
\[
 \mathfrak r_k(p)=2g(p)b(p)-\sigma_b(p).
\]
By \eqref{eq:M-KKT}, it vanishes on $0<b<1$ and is nonpositive
on $b=1$. A quadratic expansion, followed by the equal-mass
identity $\int(b^2-P)\,\dbar^3p=0$, gives the exact formula
\begin{equation}\label{eq:M-energy-identity}
 \mathcal F_{\rm M}(b)-\mathcal F_{\rm M}(P)
 =-q\int gH^2\,\dbar^3p
   +\cX(H)+q\int H\mathfrak r_k\,\dbar^3p.
\end{equation}
To check it, expand the exchange form around $P$ and then insert
$2gb=\sigma_b+\mathfrak r_k$; the mass term proportional to
$\mu_{\rm M}$ cancels. All integrals are finite by
\eqref{eq:M-H-moments} and the estimates below.

On $r<\kappa$ we have $H=0$. On $\kappa<r<k$,
$0\le g\le\delta_k^{\rm M}=O(k)$ and
$H^2\le1-b^2$, so the interior part of $q\int gH^2$ is
$O(qk^3)$. On $r>k$, $H=b$, hence
\[
 q\int gH^2\,\dbar^3p=Q_k+O(qk^3).
\]
For the residual term, $H\mathfrak r_k$ can be nonzero only
on pinned states outside the Fermi ball: on pinned states inside,
$H=0$, and on fractional states $\mathfrak r_k=0$.
On the former set, $H=1$, $0\le2g\le\sigma_b\le Ck$;
its measure is $O(k^2)$ by \eqref{eq:M-localization}. Therefore
\[
 \left|q\int H\mathfrak r_k\,\dbar^3p\right|=O(qk^3).
\]
Combining these facts with \eqref{eq:M-H-moments},
\eqref{eq:M-Q}, and \eqref{eq:M-energy-identity} yields
\[
 \cE_{\rm M}(n_k^{\rm M})-\cE_{\rm M}(P_k)
 =-\frac{qk^3}{8\pi^4}\log k+O(qk^3).
\]
Since $\rho=qk^3/(6\pi^2)$ and
$\cE_{\rm M}(P_k)=\rho e_{\rm HF}^{\rm hom}(k)$,
Theorem~\ref{thm:mueller-main} follows.

\begin{corollary}[The exchange defect is twice the leading energy gain]
\label{cor:M-defect}
For the M\"uller minimizer define the nonnegative exchange defect
per unit volume by
\[
 s_\rho=q\int b(p)(1-b(p))\sigma_b(p)\,\dbar^3p.
\]
Then
\[
 s_\rho=\frac{3}{2\pi^2}\rho\log k+O(\rho),\qquad
 \rho D_{\rm M}(k)=\frac12s_\rho+O(\rho).
\]
\end{corollary}
\begin{proof}
The constrained Euler equation implies
$s_\rho=2q\int g b^2(1-b)\,\dbar^3p$; pinned states
contribute zero. Inside the Fermi ball this is $O(qk^3)$.
Outside, subtract $2Q_k$ and use
$g b^3\le\sigma_b b^2/2$ and the exterior occupation mass
$O(k^2)$. Thus $s_\rho=2Q_k+O(qk^3)$.
\end{proof}

\begin{remark}[What the M\"uller theorem determines]
The coefficient $3/(4\pi^2)$ is the energy coefficient, not the
coefficient of $s_\rho/\rho$. The theorem has a bounded per-particle
remainder but does not identify its limit. Also, the occupied ball
$|p|<\sqrt{2\mu_{\rm M}}$ is pinned at $n=1$ at high density;
this is qualitatively different from the strictly fractional CHF
minimizer. In particular, removing the occupation cap is not an
admissible shortcut in this limit.
\end{remark}

\subsection{Symmetry reduction and the periodic thermodynamic limit}
We apply the convexity-based symmetry averaging of
\cite[Section I.C]{FLSS} to neutral periodic jellium and identify its
thermodynamic variational problem explicitly. Unlike in
Hartree--Fock theory~\cite{GHL}, restricting M\"uller jellium to
translation--spin invariant states does not change the infimum.
For $L>0$
let $\Lambda_L=\R^3/(L\mathbb Z)^3$, $\Gamma_L^*=(2\pi/L)\mathbb Z^3$,
and $\mathfrak H_L=L^2(\Lambda_L;\mathbb C^q)$. Denote by
$\mathcal A_{\rho,L}^{\rm per}$ the convex set of self-adjoint operators
$\gamma$ on $\mathfrak H_L$ satisfying $0\le\gamma\le1$,
$\operatorname{tr}\gamma=\rho L^3$, and
$\operatorname{tr}(-\Delta_L\gamma)
 :=\operatorname{tr}\bigl((-\Delta_L)^{1/2}\gamma(-\Delta_L)^{1/2}\bigr)
 <\infty$
(taking $L\to\infty$ along $\rho L^3\in\mathbb N$ if integer particle
number is imposed). With $(U_p\psi)(x)=e^{ip\cdot x}\psi(x)$, the
neutral periodic M\"uller energy with zero-mean Coulomb interaction is
\begin{equation}\label{eq:M-periodic-energy}
 \cE_{{\rm M},L}^{\rm per}(\gamma)
 =\frac12\operatorname{tr}(-\Delta_L\gamma)
  +\frac1{2L^3}\sum_{p\in\Gamma_L^*\setminus\{0\}}\frac{4\pi}{|p|^2}
   \Bigl(|\operatorname{tr}(\gamma U_p^*)|^2
   -\operatorname{tr}\bigl(\gamma^{1/2}U_p\gamma^{1/2}U_p^*\bigr)\Bigr).
\end{equation}
Adding a Madelung constant $\mu_L=c_{\rm Mad}/L$ to the periodic
potential leaves the neutral direct term unchanged and shifts the
exchange term only by the state-independent amount
$-(\mu_L/2)\operatorname{tr}\gamma=-c_{\rm Mad}\rho L^2/2$, or
$O(L^{-1})$ per particle.

\begin{proposition}[Symmetry reduction and thermodynamic limit
for periodic M\"uller jellium]
\label{prop:M-periodic}
For every $L>0$ and $\rho>0$, $\cE_{{\rm M},L}^{\rm per}$ is finite and
convex on $\mathcal A_{\rho,L}^{\rm per}$. For every
$\gamma\in\mathcal A_{\rho,L}^{\rm per}$, its translation--spin Haar
average $\overline\gamma\in\mathcal A_{\rho,L}^{\rm per}$ has the
plane-wave form
$\overline\gamma=\sum_{m\in\Gamma_L^*}n_L(m)|e_m\rangle\langle e_m|
 \otimes I_{\mathbb C^q}$ with $e_m(x)=L^{-3/2}e^{im\cdot x}$ and
satisfies
\begin{equation}\label{eq:M-averaging-gain}
 \cE_{{\rm M},L}^{\rm per}(\gamma)
 -\cE_{{\rm M},L}^{\rm per}(\overline\gamma)
 \ge\frac1{2L^3}\sum_{p\in\Gamma_L^*\setminus\{0\}}
    \frac{4\pi}{|p|^2}|\widehat{\rho_\gamma}(p)|^2\ge0,
\end{equation}
In particular, the energy is strictly decreased whenever
$\rho_\gamma$ is not constant. The infimum is attained, and at least
one minimizer is translation invariant and invariant under
$\mathrm{U}(q)$ spin rotations. Every minimizer has the constant
spatial density $\rho$; no assertion that every minimizing density
matrix is itself invariant is needed here. Moreover, with
$k=(6\pi^2\rho/q)^{1/3}$,
\begin{equation}\label{eq:M-periodic-limit}
 \lim_{L\to\infty}\frac1{\rho L^3}
 \inf_{\gamma\in\mathcal A_{\rho,L}^{\rm per}}
 \cE_{{\rm M},L}^{\rm per}(\gamma)
 =e_{\rm M}^{\rm hom}(k).
\end{equation}
The thermodynamic limit is taken at fixed $\rho$ and $q$.
The subsequent high-density limit is applied to the resulting
energy per particle.
\end{proposition}
\begin{proof}
For $\gamma\in\mathcal A_{\rho,L}^{\rm per}$, Schwarz's inequality on its
spectral decomposition gives
$\|\nabla\sqrt{\rho_\gamma}\|_{L^2(\Lambda_L)}^2
 \le\operatorname{tr}(-\Delta_L\gamma)<\infty$, so
$\rho_\gamma\in L^3(\Lambda_L)\subset L^2(\Lambda_L)$; since
$\operatorname{tr}(\gamma U_p^*)=\widehat{\rho_\gamma}(p)$ and
$|p|\ge2\pi/L$ for $p\ne0$, the direct sum is finite and convex in
$\gamma$. For each $p\ne0$,
$f_p(\gamma):=\operatorname{tr}(\gamma^{1/2}U_p\gamma^{1/2}U_p^*)
 =\|\gamma^{1/4}U_p\gamma^{1/4}\|_{\rm HS}^2\ge0$
is finite and concave on positive trace-class operators by the
Wigner--Yanase theorem, as in~\cite{FLSS}. For spatial translations $T_a$
($a\in\Lambda_L$) and spin rotations $R_U$ ($U\in\mathrm{U}(q)$), put
$V_{a,U}=T_aR_U$. Then $V_{a,U}^*U_pV_{a,U}=e^{-ip\cdot a}U_p$ and hence
$f_p(V_{a,U}\gamma V_{a,U}^*)=f_p(\gamma)$.

The Haar Bochner average
$\overline\gamma=L^{-3}\int_{\Lambda_L}\mathrm{d}^3a
 \int_{\mathrm{U}(q)}\mathrm{d}\mu(U)\,V_{a,U}\gamma V_{a,U}^*$
lies in $\mathcal A_{\rho,L}^{\rm per}$, satisfies
$\operatorname{tr}(-\Delta_L\overline\gamma)
 =\operatorname{tr}(-\Delta_L\gamma)$ and
$\widehat{\rho_{\overline\gamma}}(p)=0$ for $p\ne0$, and has the stated
plane-wave form by Schur's lemma, with $0\le n_L(m)\le1$ and
$(q/L^3)\sum_m n_L(m)=\rho$. Approximate $\overline\gamma$ in trace norm
by Riemann-sum convex combinations $\gamma_j$ of $V_{a,U}\gamma V_{a,U}^*$.
Since $\|\gamma_j^{1/2}-\overline\gamma^{1/2}\|\to0$ and
$\operatorname{tr}\gamma_j=\rho L^3=\operatorname{tr}\overline\gamma$,
testing against finite-rank spectral projections of $\overline\gamma$
yields $\operatorname{tr}(\gamma_j^{1/2}\overline\gamma^{1/2})
 \to\operatorname{tr}\overline\gamma$ and hence
$\|\gamma_j^{1/2}-\overline\gamma^{1/2}\|_{\rm HS}\to0$. Thus
$f_p(\gamma_j)=\langle U_p\gamma_j^{1/2},\gamma_j^{1/2}U_p\rangle_{\rm HS}
 \to f_p(\overline\gamma)$, while concavity gives
$f_p(\gamma_j)\ge f_p(\gamma)$, so $f_p(\overline\gamma)\ge f_p(\gamma)$
for every $p\ne0$.

Associate to $n_L$ the step function $n^{(L)}\in\mathcal A_\rho$ equal to
$n_L(m)$ on $Q_m=m+[-\pi/L,\pi/L)^3$. Its kinetic moment is
\[
 \int |p|^2n^{(L)}(p)\,\dbar^3p
 =L^{-3}\sum_{m\in\Gamma_L^*}|m|^2n_L(m)
  +\frac{\pi^2\rho}{qL^2}.
\]
To make the exchange comparison explicit, set
$Q_0^{\rm unit}=[-1/2,1/2)^3$ and, for $\nu\in\mathbb Z^3$, define
\[
 a_\nu=\iint_{Q_0^{\rm unit}\times Q_0^{\rm unit}}
       \frac{\mathrm{d}^3u\,\mathrm{d}^3v}{|\nu+u-v|^2},
 \qquad
 c_\nu=
 \begin{cases}
  |\nu|^{-2},&\nu\ne0,\\
  0,&\nu=0.
 \end{cases}
\]
The singularity is locally integrable, so each $a_\nu$ is finite,
including $a_0$. For large $|\nu|$, the linear Taylor term vanishes
by symmetry and
$a_\nu=|\nu|^{-2}+O(|\nu|^{-4})$. Thus
\[
 \sum_{\nu\in\mathbb Z^3}|a_\nu-c_\nu|<\infty.
\]
Using $f_p(\overline\gamma)=q\sum_m\sqrt{n_L(m)n_L(m-p)}$ and
\[
 \sum_{m\in\Gamma_L^*}
 \sqrt{n_L(m)n_L(m-2\pi\nu/L)}
 \le\sum_m n_L(m)=\frac{\rho L^3}{q},
\]
rescaling the two momentum cells gives
\begin{equation}\label{eq:M-cell-exchange}
 \left|\frac1{2L^6}\sum_{p\in\Gamma_L^*\setminus\{0\}}
 \frac{4\pi}{|p|^2}f_p(\overline\gamma)
 -\cX\bigl(\sqrt{n^{(L)}}\bigr)\right|
 \le\frac{C\rho}{L},
\end{equation}
where $C$ is absolute. By \eqref{eq:dual-hardy},
$\cX(\sqrt{n^{(L)}})<\infty$; since $0\le f_p(\gamma)\le f_p(\overline\gamma)$,
the exchange sum in \eqref{eq:M-periodic-energy} converges for every
$\gamma\in\mathcal A_{\rho,L}^{\rm per}$, proving finiteness and
convexity. More precisely, the averaging argument gives
\[
\begin{aligned}
 &\cE_{{\rm M},L}^{\rm per}(\gamma)
   -\cE_{{\rm M},L}^{\rm per}(\overline\gamma)\\
 &\quad=D_L(\rho_\gamma-\rho)
  +\frac1{2L^3}\sum_{p\in\Gamma_L^*\setminus\{0\}}
   \frac{4\pi}{|p|^2}
   \bigl[f_p(\overline\gamma)-f_p(\gamma)\bigr],
\end{aligned}
\]
where $D_L(\rho_\gamma-\rho)$ denotes the nonnegative direct term in
\eqref{eq:M-periodic-energy}. This proves
\eqref{eq:M-averaging-gain}. The kinetic identity and
\eqref{eq:M-cell-exchange} also yield
\[
 \frac{\cE_{{\rm M},L}^{\rm per}(\overline\gamma)}{L^3}
 =\cE_{\rm M}(n^{(L)})-\frac{\pi^2\rho}{2L^2}
   +O(\rho/L),
\]
and hence the lower bound
\[
 (\rho L^3)^{-1}\cE_{{\rm M},L}^{\rm per}(\gamma)
 \ge e_{\rm M}^{\rm hom}(k)-O(L^{-1}+L^{-2}).
\]

We next prove attainment for fixed $L$ and $N=\rho L^3$.
Let $v_L$ be the zero-mean periodic Coulomb potential from
\eqref{eq:M-periodic-energy}, and let $X_L(A)$ denote its exchange
quadratic form in the kernel $A$. The bound
\[
 |v_L(z)|\le
 \frac{C}{\operatorname{dist}_{\Lambda_L}(z,0)}+\frac{C}{L}
\]
and the periodic Hardy inequality, applied in the first kernel
variable and then integrated in the second, imply
\[
 |X_L(\sqrt\gamma)|
 \le C\sqrt{N\,T_L(\gamma)}+C\frac NL,
 \qquad
 T_L(\gamma)=\operatorname{tr}(-\Delta_L\gamma).
\]
Here $\|\sqrt\gamma\|_{\rm HS}^2=N$ and
$\|\nabla_x\sqrt\gamma\|_{\rm HS}^2=T_L(\gamma)$.
Since the direct term is nonnegative, a minimizing sequence
$\gamma_j$ has uniformly bounded kinetic energy. The compact
resolvent of $-\Delta_L$ then gives, after extraction,
$\gamma_j\to\gamma$ in trace norm. Indeed, the trace outside a
kinetic spectral cutoff at energy $M$ is bounded by
$T_L(\gamma_j)/M$; the corresponding finite-rank compressions
are compact in trace norm. Thus $0\le\gamma\le1$,
$\operatorname{tr}\gamma=N$, and
$T_L(\gamma)\le\liminf_jT_L(\gamma_j)$.

Put $A_j=\sqrt{\gamma_j}$ and $A=\sqrt\gamma$.
The square-root continuity used above gives
$\|A_j-A\|_{\rm HS}\to0$. The periodic Hardy inequality and
Schwarz's inequality yield
\[
\begin{aligned}
 |X_L(A_j)-X_L(A)|
 \le{}&C\|A_j-A\|_{\rm HS}\\
 &\times\left(
 \|\nabla_xA_j\|_{\rm HS}+\|\nabla_xA\|_{\rm HS}
 +L^{-1}\sqrt N\right)\longrightarrow0.
\end{aligned}
\]
The direct term is lower semicontinuous by its nonnegative Fourier
series: each coefficient is continuous under trace-norm convergence,
and Fatou's lemma applies to the sum. Hence $\gamma$ attains the
infimum. Averaging a minimizer provides a translation--spin invariant
minimizer. Moreover, \eqref{eq:M-averaging-gain} forces every minimizer
to have constant spatial density. This does not require uniqueness
of the minimizing density matrix.

For the upper thermodynamic bound, let $n\in\mathcal A_\rho$
minimize $\cE_{\rm M}$ and set
$n_L(m)=L^3\int_{Q_m}n\,\dbar^3p$. For $p\in Q_m$,
$|m|^2\le|p|^2+2\sqrt3\pi L^{-1}|p|+3\pi^2L^{-2}$, so the associated
plane-wave state $\overline\gamma_L\in\mathcal A_{\rho,L}^{\rm per}$
satisfies $(2L^3)^{-1}\operatorname{tr}(-\Delta_L\overline\gamma_L)
 \le\frac q2\int|p|^2n\,\dbar^3p+O(L^{-1})$. Moreover, the step function
$n^{(L)}$ converges to $n$ in $L^1(\R^3,\dbar^3p)$, hence
$\|\sqrt{n^{(L)}}-\sqrt n\|_2^2\le\|n^{(L)}-n\|_1\to0$ with uniformly
bounded kinetic moment. Lemma~\ref{lem:compact} and
\eqref{eq:M-cell-exchange} give
$\limsup_{L\to\infty}(\rho L^3)^{-1}
 \cE_{{\rm M},L}^{\rm per}(\overline\gamma_L)\le e_{\rm M}^{\rm hom}(k)$,
proving \eqref{eq:M-periodic-limit}.
\end{proof}

\section{The CHF minimizer: strict occupation and uniqueness}\label{sec:CHF-structure}
\subsection{A strict competitor for the Fermi ball}
Use the volume coordinate \eqref{eq:volume}.
If $n\in\mathcal A_\rho$ satisfies $w_n=0$ almost everywhere,
then $\cE(n)=\cE_{\rm HF}(n)$. The bathtub principle and
\eqref{eq:exchange-maximality} therefore imply
\begin{equation}\label{eq:indicator}
 \cE(n)\ge\cE(P_k).
\end{equation}
Here $w_n=0$ is equivalent to $n\in\{0,1\}$ almost everywhere.

\begin{lemma}[The Fermi sea is not optimal]\label{lem:strict}
For every $\rho>0$, $\inf\cE<\cE(P_k)$.
\end{lemma}
\begin{proof}
For $0<\varepsilon<1/2$ and $0<\delta<k/6$, define
\[
 n_{\varepsilon,\delta}(t)=
 \begin{cases}
 1-\varepsilon,&-\delta<t<0,\\
 \varepsilon,&0<t<\delta,\\
 \one_{\{t<0\}},&|t|\ge\delta.
 \end{cases}
\]
It has exactly the required mass, and
$w_{n_{\varepsilon,\delta}}=
\sqrt{\varepsilon(1-\varepsilon)}\one_{\{|t|<\delta\}}$.
Using \eqref{eq:radial-X} and \eqref{eq:fermi-field} and expanding at $t=0$ gives,
for fixed $k,\varepsilon$, as $\delta\downarrow0$,
\begin{align*}
 \cE(n_{\varepsilon,\delta})-\cE(P_k)
 \le \frac{q\varepsilon k^2\delta^2}{2\pi^3}
 \left[\pi k-(1-2\varepsilon)\log\frac{k}{\delta}
       +O_{k,\varepsilon}(1)\right].
\end{align*}
The quadratic contribution $-\cX(n_{\varepsilon,\delta}-P_k)$
may be dropped for this upper bound. Since $1-2\varepsilon>0$,
the right-hand side is negative when $\delta$ is small.
\end{proof}

\begin{lemma}[Strictly fractional occupations]\label{lem:fractional}
Every minimizer $n_*$ satisfies $0<n_*<1$ almost everywhere.
\end{lemma}
\begin{proof}
Lemma~\ref{lem:strict} and \eqref{eq:indicator} show that
$w_*:=\sqrt{n_*(1-n_*)}$ is not identically zero.
Consequently $\Delta_*:=Vw_*>0$ at every finite momentum; it is
bounded and continuous because $w_*\le1/2$ and $w_*\in L^2$.
If $\{n_*=0\}$ has positive measure, choose a bounded
positive-measure subset $A$ of that set, and a bounded set $B$ of
positive measure on which $\eta\le n_*\le1-\eta$ for some $\eta>0$.
For small $\varepsilon>0$ the mass-preserving variation
\[
 n_\varepsilon=n_*+\varepsilon\one_A
 -\varepsilon\frac{|A|}{|B|}\one_B
\]
is admissible. Kinetic energy and ordinary exchange change by
$O(\varepsilon)$. On $A$ the new $w$ is
$\sqrt\varepsilon+O(\varepsilon^{3/2})$, while its change on $B$ is
$O(\varepsilon)$. Since $\cX$ is a positive quadratic form,
\[
 \cX(w_{n_\varepsilon})-\cX(w_*)
 \ge q\sqrt\varepsilon\int_A\Delta_*\,\dbar^3p+O(\varepsilon).
\]
The leading coefficient is positive, contradicting minimality.
The set $\{n_*=1\}$ is excluded by transferring mass in the opposite
direction.
\end{proof}

\begin{theorem}[Uniqueness and self-consistency]\label{thm:unique}
The minimizer $n_\rho$ is unique. It has a continuous radial
representative satisfying $0<n_\rho(p)<1$ for every $p$.
There is a unique $\mu_\rho\in\R$ such that, with
\begin{equation}\label{eq:gap-notation}
 \xi(p)=\frac{|p|^2}{2}-(Vn_\rho)(p)-\mu_\rho,
 \quad\Delta(p)=(Vw_\rho)(p),\quad
 E(p)=\sqrt{\xi(p)^2+\Delta(p)^2},
\end{equation}
one has pointwise
\begin{equation}\label{eq:gap}
 n_\rho=\frac12\left(1-\frac\xi E\right),\qquad
 w_\rho=\frac\Delta{2E},\qquad \Delta>0.
\end{equation}
\end{theorem}
\begin{proof}
Local variations on
$\{p:|p|\le j,\ j^{-1}\le n_\rho(p)\le1-j^{-1}\}$,
with zero integral, give the common Lagrange multiplier $\mu_\rho$
and the a.e. Euler equation
\begin{equation}\label{eq:EL}
 \frac{|p|^2}{2}-(Vn_\rho)(p)-\mu_\rho
 =\Delta(p)\frac{1-2n_\rho(p)}{2w_\rho(p)}.
\end{equation}
The fields are continuous, bounded, and $\Delta>0$. Algebraic
inversion of \eqref{eq:EL} gives \eqref{eq:gap}, defining a
continuous representative strictly between zero and one.
Uniqueness of $\mu_\rho$ is immediate from \eqref{eq:EL}.

Let $n$ be another admissible occupation of the same mass and put
$h=n-n_\rho$, $g=w_n-w_\rho$. Since $\int h\,\dbar^3p=0$,
expansion of the quadratic forms gives
\begin{equation}\label{eq:quadratic-stability}
 \cE(n)-\cE(n_\rho)
 =q\int E(h^2+g^2)\,\dbar^3p-\cX(h)-\cX(g).
\end{equation}
Here one uses the exact pointwise circle identity
\[
 h^2+g^2=(1-2n_\rho)h-2w_\rho g
\]
and the Euler relations $\xi=E(1-2n_\rho)$, $\Delta=2Ew_\rho$.
We justify symmetrization and the use of Fubini's theorem. The bounded
self-consistent potentials imply $0<E(p)\le C(1+|p|^2)$, with $\rho$
fixed. For $f=h$ or $f=g$, the inequalities
\[
 h^2\le n+n_\rho,\qquad g^2\le2(n+n_\rho)
\]
and the finite kinetic energies give
\begin{equation}\label{eq:Ef-integrable}
 \int_{\R^3}E(p)f(p)^2\,\dbar^3p<\infty.
\end{equation}
Put $K(p,p')=4\pi|p-p'|^{-2}$. Since $w_\rho>0$ and
$Vw_\rho=2Ew_\rho$, Tonelli's theorem gives
\[
\begin{aligned}
 &\iint K(p,p')w_\rho(p)w_\rho(p')
   \frac{f(p)^2}{w_\rho(p)^2}\,\dbar^3p\,\dbar^3p'\\
 &\qquad=\int\frac{f(p)^2}{w_\rho(p)}(Vw_\rho)(p)\,\dbar^3p
 =2\int E(p)f(p)^2\,\dbar^3p<\infty.
\end{aligned}
\]
The same holds with $p$ and $p'$ interchanged. Moreover,
\[
 2\left|\frac{f(p)f(p')}{w_\rho(p)w_\rho(p')}\right|
 \le \frac{f(p)^2}{w_\rho(p)^2}
     +\frac{f(p')^2}{w_\rho(p')^2}.
\]
All terms in the expansion of the squared difference are therefore
absolutely integrable. Fubini's theorem applies and yields
\[
\begin{aligned}
 &\iint K(p,p')w_\rho(p)w_\rho(p')
 \left(\frac{f(p)}{w_\rho(p)}-\frac{f(p')}{w_\rho(p')}\right)^2
 \,\dbar^3p\,\dbar^3p'\\
 &\qquad=4\int E(p)f(p)^2\,\dbar^3p
          -2\int f(p)(Vf)(p)\,\dbar^3p.
\end{aligned}
\]
Applying this identity to $h$ and $g$ gives
\begin{equation}\label{eq:stability}
\begin{split}
\cE(n)-\cE(n_\rho)
=\frac q4\iint \frac{4\pi w_\rho(p)w_\rho(p')}{|p-p'|^2}
\bigg[&\left(\frac{h(p)}{w_\rho(p)}
             -\frac{h(p')}{w_\rho(p')}\right)^2\\
     +&\left(\frac{g(p)}{w_\rho(p)}
             -\frac{g(p')}{w_\rho(p')}\right)^2\bigg]
\,\dbar^3p\,\dbar^3p'.
\end{split}
\end{equation}
The right-hand side is finite and nonnegative. If it vanishes, then
$h=cw_\rho$ almost everywhere, since the kernel is strictly positive.
Here $h\in L^1$ and $\int h\,\dbar^3p=0$; positivity of $w_\rho$
therefore forces $c=0$. This does not require $w_\rho\in L^1$:
if $c\ne0$, then $h=cw_\rho$ has a fixed nonzero sign, contradicting
its zero integral. Hence every minimizer equals $n_\rho$.
Rotations preserve the problem; uniqueness gives radiality.
\end{proof}

For the connection between CHF optimization and Hartree--Fock--Bogoliubov
methods, see Staroverov and Scuseria~\cite{SS}. The self-consistent
equations and the stability identity needed here have been derived
directly above.

\section{A coercive deterministic comparison for CHF}\label{sec:direct}
The upper bound will follow from a comparison for every
$n\in\mathcal A_\rho$, without using its Euler equation or radiality.
Put
\begin{align}
 \mu_{\rm HF}&=\frac{k^2}{2}-\frac{k}{\pi},\notag\\
 \xi_{\rm HF}(r)
 &=\frac{r^2}{2}-(VP_k)(r)-\mu_{\rm HF}\notag\\
 &=\frac{r^2-k^2}{2}
 \left(1+\frac1{\pi r}\log\frac{r+k}{|r-k|}\right),
 \label{eq:xi-HF}\\
 e_k(r)&:=|\xi_{\rm HF}(r)|\ge\frac12|r^2-k^2|.
 \label{eq:eglob}
\end{align}
The limiting values at $r=0,k$ are understood by continuity.
In particular $\xi_{\rm HF}$ has the sign of $r-k$.
For $h=n-P_k$, $d=|h|$, and $s_k=1-2P_k$ we have
$h=s_kd$ and $w_n^2=d-d^2$. Expanding ordinary exchange and using
$\int h\,\dbar^3p=0$ gives the exact identity
\begin{equation}\label{eq:energy-start}
 \boxed{\cE(n)-\cE(P_k)
 =q\int e_kd\,\dbar^3p-\cX(h)-\cX(w_n).}
\end{equation}

For $R=\theta k$ and $\Omega=\{p:|t(|p|)|\le R\}$, choose $a>0$ and set
\begin{equation}\label{eq:Ea}
 E_a=\sqrt{e_k^2+k^2a^2\one_\Omega},\qquad
 d_a=\frac12\left(1-\frac{e_k}{E_a}\right),\qquad
 \|f\|_{E_a}^2=\int E_a|f|^2\,\dbar^3p.
\end{equation}
Thus $0\le d_a\le1/2$ and $d_a=0$ outside the shell. This is a
comparison defect, not a trial occupation; no particle-number
constraint is imposed on $d_a$.
Define
\begin{equation}\label{eq:Ba}
 \mathscr B_a=\tfrac12 E_a^{-1/2}VE_a^{-1/2},\qquad
 \beta(a)=\|\mathscr B_a\|,\qquad
 \beta_\pm(a)=\|\Pi_\pm\mathscr B_a\Pi_\pm\|,
\end{equation}
where $\Pi_-=P_k$, $\Pi_+=1-P_k$ are multiplication operators on
$L^2(\R^3,\dbar^3p)$. The next section constructs a deterministic
$a$ for which the following hypotheses hold.

\begin{proposition}[Coercive comparison]\label{prop:completion}
If $\beta(a)\le1$ and $\beta_\pm(a)\le3/5$, then every
$n\in\mathcal A_\rho$ satisfies
\begin{equation}\label{eq:coercive-main}
 \boxed{\cE(n)-\cE(P_k)
 \ge -q\|d_a\|_{E_a}^2-5\cX(d_a)
       +\frac q4\bigl\||n-P_k|-d_a\bigr\|_{E_a}^2.}
\end{equation}
No symmetry or minimizing property of $n$ is assumed.
\end{proposition}
\begin{proof}
The full norm bound gives $\cX(w_n)\le q\int E_aw_n^2\,\dbar^3p$.
Substitute $w_n^2=d-d^2$ into \eqref{eq:energy-start} and complete
the square:
\begin{equation}\label{eq:complete-raw}
 \cE(n)-\cE(P_k)
 \ge q\|d-d_a\|_{E_a}^2-q\|d_a\|_{E_a}^2-\cX(h).
\end{equation}
We must still control the last negative term. If $f\le0$ inside
the Fermi ball and $f\ge0$ outside, its mixed Coulomb contribution
between the two regions is nonpositive. The compressed norms give
\begin{equation}\label{eq:sign-cone}
 \cX(f)\le\cX(\Pi_-f)+\cX(\Pi_+f)
 \le\tfrac35q\|f\|_{E_a}^2.
\end{equation}
Apply this to the positive-part decomposition
\[
 h=h_1+h_0,\qquad
 h_1=s_k(d-d_a)_+,\qquad h_0=s_k\min\{d,d_a\}.
\]
Here $h_1$ has the required sign and $|h_0|\le d_a$, so positivity
of the kernel implies $\cX(h_0)\le\cX(d_a)$.
Schwarz for the positive Coulomb form and
$2uv\le u^2/4+4v^2$ imply
\[
\begin{aligned}
 \cX(h)&\le\tfrac54\cX(h_1)+5\cX(h_0)\\
 &\le\tfrac34q\|(d-d_a)_+\|_{E_a}^2+5\cX(d_a)
 \le\tfrac34q\|d-d_a\|_{E_a}^2+5\cX(d_a).
\end{aligned}
\]
Insert this into \eqref{eq:complete-raw}. All expressions are finite:
$E_a(p)\le C_{k,a}(1+|p|^2)$, the competitors have finite kinetic
energy, and $d_a$ is bounded with compact support.
\end{proof}

The full norm bound controls the pairing exchange, whereas the
one-sided bounds control ordinary exchange without changing the
leading source term $q\|d_a\|_{E_a}^2$. The numerical constants
$3/5$, $5$, and $1/4$ only provide a fixed coercive margin; they
will not enter the asymptotic energy coefficient.

\section{The deterministic spectral threshold}\label{sec:det-spectrum}
\subsection{The logarithmic energy space and model}
Put $L=\log(R/a)$ and $\ell=\log(2/\theta)$. The model weight is
\begin{equation}\label{eq:modweight}
 W_{L,k}(x)=\left[\left(1+\frac{x+\ell}{\pi k}\right)^2
                        +e^{2(x-L)}\right]^{-1/2}.
\end{equation}
For $\ell>0$ let
\[
 \mathcal H_\ell=\{u\in\mathrm{AC}_{\rm loc}([0,\infty)):
                        u'\in L^2(0,\infty)\},
\]
with inner product and squared norm
\[
 \mathfrak a_\ell[u,v]
 =\int_0^\infty u'\overline{v'}\,\mathrm{d}x
   +\ell^{-1}u(0)\overline{v(0)},\qquad
 \mathfrak a_\ell[u]=\mathfrak a_\ell[u,u].
\]
The map $u\mapsto(\ell^{-1/2}u(0),u')$ is an isometric isomorphism
onto $\mathbb C\oplus L^2(0,\infty)$. No $L^2$ condition is imposed
on $u$ itself. The reproducing kernel is
$G_\ell(x,y)=\ell+\min(x,y)$, since
$\mathfrak a_\ell[u,G_\ell(\cdot,y)]=u(y)$. In particular,
\begin{equation}\label{eq:energy-evaluation}
 |u(x)|^2\le(x+\ell)\mathfrak a_\ell[u].
\end{equation}
For $W\ge0$ with $\int_0^\infty(x+\ell)W(x)\,\mathrm{d}x<\infty$, put
\begin{equation}\label{eq:Lambda}
 \Lambda_\ell(W)=\sup_{0\ne u\in\mathcal H_\ell}
 \frac{\int_0^\infty W|u|^2\,\mathrm{d}x}{\mathfrak a_\ell[u]}.
\end{equation}
The associated compact positive operator $S_W$ on $\mathcal H_\ell$
is defined by $\mathfrak a_\ell[S_Wu,v]=\int Wu\overline v$.
It has norm $\Lambda_\ell(W)$; \eqref{eq:energy-evaluation} also gives
\begin{equation}\label{eq:weight-stability}
 \|S_{W_1}-S_{W_2}\|
 \le\int_0^\infty(x+\ell)|W_1-W_2|\,\mathrm{d}x.
\end{equation}
Compactness follows, for example, because the embedding
$u\mapsto\sqrt W u$ is Hilbert--Schmidt, with squared norm
$\int(x+\ell)W$.

\begin{proposition}[Model eigenvalue]\label{prop:modelexp}
Fix $\ell>0$ and constants $0<c_0\le C_0<\infty$.
As $L\to\infty$, uniformly for $k>0$ satisfying
$c_0\le L^2/k\le C_0$, we have
\begin{equation}\label{eq:model-spectrum}
 \Lambda_\ell(W_{L,k})=
 \frac{4L^2}{\pi^2}+\frac{8(\ell+\log2)L}{\pi^2}
 -\frac{4L^3}{\pi^3k}\left(\frac12+\frac2{\pi^2}\right)
 +O_{\ell,c_0,C_0}(1).
\end{equation}
\end{proposition}
\begin{proof}
In this proof, constants may also depend on $c_0,C_0$;
this dependence is suppressed in the notation.
For $W_L^0=\one_{[0,L]}$, the first eigenvalue is
$\lambda_0=\omega_0^{-2}$, where $\cot(\omega_0L)=\ell\omega_0$.
Thus
\begin{equation}\label{eq:sharp-model}
 \lambda_0=\frac4{\pi^2}(L+\ell)^2+O_\ell(1),
\end{equation}
and the gap to the next eigenvalue is of order $L^2$.
The eigenfunction normalized by $\mathfrak a_\ell[u_0]=1$ is
$c_L\cos(\omega_0(L-x))$ on $[0,L]$, continued constantly after $L$.
Integration by parts gives
\[
 1=\omega_0^2\int_0^L u_0^2\,\mathrm{d}x,\qquad
 c_L^2=\frac{2}{\omega_0^2
        (L+\ell/(1+\ell^2\omega_0^2))}.
\]
Since $\omega_0=\pi/(2L)+O_\ell(L^{-2})$ and
$c_L^2=8L/\pi^2+O_\ell(1)$, uniformly on $[0,L]$,
\begin{equation}\label{eq:u0-uniform}
 \sup_{0\le x\le L}\left|
 u_0(x)^2-\frac{8L}{\pi^2}\sin^2\frac{\pi x}{2L}\right|
 \le C_\ell.
\end{equation}
Indeed the phases differ by $O_\ell(L^{-1})$, and the squared cosine
is Lipschitz. Also $u_0(L)^2=8L/\pi^2+O_\ell(1)$ and
$|(u_0^2)'|\le C_\ell$.

Let $g(z)=(1+e^{2z})^{-1/2}$. Taylor expansion in
$(x+\ell)/(\pi k)$ gives
\[
 W_{L,k}=W_L^0+[g(x-L)-W_L^0]
           -\frac{x+\ell}{\pi k}W_L^0+r_{L,k},
\]
where
\begin{equation}\label{eq:model-weight-remainder}
 \int_0^\infty(x+\ell)|r_{L,k}(x)|\,\mathrm{d}x
 \le C_\ell\left(\frac{1+L^4}{k^2}+\frac{1+L^2}{k}\right)
 =O_\ell(1).
\end{equation}
To see this, the quadratic Taylor remainder is bounded by
$C((x+\ell)/k)^2(1+e^{2(x-L)})^{-3/2}$.
In the linear term replace $(1+e^{2(x-L)})^{-3/2}$ by $W_L^0$;
the difference has finite absolute moments about $x=L$, and
contributes $O_\ell((1+L^2)/k)$.

The total perturbation norm in \eqref{eq:weight-stability} is $O(L)$.
The spectral gap therefore makes the remainder after first-order
eigenvalue perturbation $O(L^2/L^2)=O(1)$.
The transition contributes
$8(\log2)L/\pi^2+O_\ell(1)$, because
\begin{equation}\label{eq:logtwo}
 \int_\R[g(z)-\one_{\{z<0\}}]\,\mathrm{d}z=\log2
\end{equation}
and its first absolute moment is finite. The bound on $(u_0^2)'$
justifies replacing $u_0(L+z)^2$ by $u_0(L)^2$ at a cost $O(1)$.
Finally, \eqref{eq:u0-uniform} gives
\[
 -\frac1{\pi k}\int_0^L(x+\ell)u_0(x)^2\,\mathrm{d}x
 =-\frac{4L^3}{\pi^3k}\left(\frac12+\frac2{\pi^2}\right)+O_\ell(1),
\]
using
\begin{equation}\label{eq:sinmoment}
 \int_0^L x\sin^2\frac{\pi x}{2L}\,\mathrm{d}x
 =L^2\left(\frac14+\frac1{\pi^2}\right).
\end{equation}
The uniform $O_\ell(1)$ error in \eqref{eq:u0-uniform} contributes
only $O_\ell(L^2/k)=O_\ell(1)$. Adding the Robin term in
\eqref{eq:sharp-model} proves \eqref{eq:model-spectrum}.
\end{proof}

\subsection{Full and one-sided Coulomb norms}
For the radial weight $E_a$, spherical-harmonic decomposition and
$|P_j(z)|\le1$ show that the full norm and both compressed norms in
\eqref{eq:Ba} equal their norms in the nonnegative radial channel.
The radial kernel on $L^2(\R_+,\mathrm{d}r)$ is
\begin{equation}\label{eq:Bradial}
 \mathscr B_a(r,s)=\frac1{2\pi}
 \frac{\log((r+s)/|r-s|)}{\sqrt{E_a(r)E_a(s)}}.
\end{equation}
It is Hilbert--Schmidt: the diagonal logarithm is locally square
integrable, and $E_a(r)\asymp r^2$ at infinity for fixed $k,a$.
Angular domination gives the corresponding bounded full-space forms.
The degenerate-dispersion Coulomb analysis is related to the approach
of Gontier, Hainzl, and Lewin~\cite{GHL}, whose variational problem
allows translation-symmetry breaking in Hartree--Fock theory; no
such enlargement of the present class is made here.

\begin{lemma}[Deterministic spectral comparison]\label{lem:det-spectrum}
Fix $0<\theta<1/12$ and constants
$0<c_0\le C_0<\infty$. As $L=\log(R/a)\to\infty$,
uniformly for $k>0$ satisfying $c_0\le L^2/k\le C_0$,
\begin{align}
 \pi k\beta(a)&=\Lambda_\ell(W_{L,k})
                   +O_{\theta,c_0,C_0}(1+\sqrt L),
 \label{eq:full-spectrum}\\
 2\pi k\beta_\pm(a)&=\Lambda_\ell(W_{L,k})
                   +O_{\theta,c_0,C_0}(1+\sqrt L).
 \label{eq:half-spectrum}
\end{align}
\end{lemma}
\begin{proof}
In this proof, dependence of constants on $c_0,C_0$
is again suppressed.
On the two sides of the shell use
$t_\sigma(x)=\sigma Re^{-x}$, $r_\sigma(x)=r(t_\sigma(x))$,
$x\ge0$, and set
\[
 \omega_{a,\sigma}(x)=\frac{kRe^{-x}}{E_a(r_\sigma(x))}.
\]
The exact Jacobian is
$|\mathrm{d}r_\sigma/\mathrm{d}x|=k^2Re^{-x}/r_\sigma^2$.
The shell block of $2\pi k\mathscr B_a$ consequently has matrix kernel
\[
 \sqrt{\omega_{a,\sigma}(x)\omega_{a,\tau}(y)}
 \frac{k^2}{r_\sigma(x)r_\tau(y)}
 \log\frac{r_\sigma(x)+r_\tau(y)}{|r_\sigma(x)-r_\tau(y)|}.
\]
The cubic-coordinate identity gives
\begin{align}
 \log\frac{r_\sigma(x)+r_\tau(y)}{|r_\sigma(x)-r_\tau(y)|}
 &=\min(x,y)+\ell
 -\log|1-\sigma\tau e^{-|x-y|}|+G_{\sigma\tau}(x,y),
 \label{eq:log-kernel-decomp}\\
 |G_{\sigma\tau}(x,y)|+
 \left|\frac{k^2}{r_\sigma(x)r_\tau(y)}-1\right|
 &\le C_\theta(e^{-x}+e^{-y}).
 \label{eq:geom-operator}
\end{align}
The explicit dispersion \eqref{eq:xi-HF} implies
\[
 \frac{e_k(r_\sigma(x))}{kRe^{-x}}
 =1+\frac{x+\ell}{\pi k}+\epsilon_\sigma(x),\qquad
 |\epsilon_\sigma(x)|
 \le C\theta e^{-x}\left(1+\frac{x+\ell}{k}\right).
\]
The exact and model dispersion factors are positive and uniformly
comparable. With $b=e^{x-L}$, the derivative of
$F(A)=(A^2+b^2)^{-1/2}$ satisfies
$|F'(A)|\le C_\theta/(1+b^2)$ along their connecting segment.
Hence
\begin{equation}\label{eq:det-weight-error}
 \int_0^\infty(x+\ell)|\omega_{a,\sigma}-W_{L,k}|\,\mathrm{d}x
 \le C_\theta.
\end{equation}
Also
\begin{equation}\label{eq:det-envelope}
 \omega_{a,\sigma}\le C_\theta\min\{1,e^{L-x}\},\quad
 \int_0^\infty\omega_{a,\sigma}=O_\theta(1+L),\quad
 \int_0^\infty e^{-2x}(x+\ell)^2\omega_{a,\sigma}\,\mathrm{d}x
 \le C_\theta.
\end{equation}
The functions $-\log|1\mp e^{-|z|}|$ belong to $L^1(\R)$.
Their weighted convolution kernels have norm $O_\theta(1)$ by
Schur's test. For the geometric terms use
$\min(x,y)+\ell\le x+\ell$ when accompanied by $e^{-x}$,
and symmetrically for $e^{-y}$. Their Hilbert--Schmidt norms are
$O_\theta(1+\sqrt L)$ by \eqref{eq:det-envelope}.

For the leading matrix define
$J:\mathcal H_\ell\to L^2(0,\infty)\oplus L^2(0,\infty)$ by
$Ju=(\sqrt{\omega_{a,+}}u,\sqrt{\omega_{a,-}}u)$.
It is Hilbert--Schmidt, with squared norm
$\sum_\sigma\int(x+\ell)\omega_{a,\sigma}$.
The reproducing identity on $\mathcal H_\ell$ shows that $JJ^*$
has exactly the leading matrix kernel. Its norm is
$2\Lambda_\ell((\omega_{a,+}+\omega_{a,-})/2)$;
for a single side it is $\Lambda_\ell(\omega_{a,\sigma})$.
The adjoints here use $\mathfrak a_\ell$ on the domain of $J$.
Equations \eqref{eq:weight-stability} and \eqref{eq:det-weight-error}
therefore yield the required shell estimates, including the factor two.

Outside the shell, \eqref{eq:eglob} implies
$E_a(r)=e_k(r)\ge c_\theta(k^2+r^2)$.
After $r=kz$, the exterior--exterior block of $2\pi k\mathscr B_a$
has Hilbert--Schmidt norm $O_\theta(1)$.
For the mixed block use
\[
 \sup_{z\in I_\theta}\int_{I_\theta^c}
 \frac{\log^2((z+z')/|z-z'|)}{1+z'^2}\,\mathrm{d}z'<\infty,
 \qquad
 I_\theta=[(1-3\theta)^{1/3},(1+3\theta)^{1/3}].
\]
The logarithmic singularity at a shell endpoint is square integrable,
and the integrand decays at infinity. Integrating the remaining shell
factor with \eqref{eq:det-envelope} gives norm $O_\theta(1+\sqrt L)$.
The same estimates hold for the one-sided restrictions.
Restoring the factor $1/(2\pi k)$ proves both assertions.
\end{proof}

\subsection{An explicit subcritical comparison scale}
Define, using \eqref{eq:CHF-constants},
\begin{equation}\label{eq:chosen-scale}
 a_0(k)=y_0k e^{-A\sqrt{k}},\qquad
 a_*(k)=a_0(k)e^{k^{-1/8}},\qquad
 E_*=E_{a_*},\quad d_*=d_{a_*}.
\end{equation}
The vanishing safety margin makes the full norm strictly subcritical
without changing the leading constant.
\begin{corollary}\label{cor:subcritical}
For all sufficiently large $k$,
\begin{equation}\label{eq:norms}
 \beta(a_*)\le1,\qquad \beta_\pm(a_*)\le\frac35,
 \qquad\beta_\pm(a_*)\longrightarrow\frac12.
\end{equation}
\end{corollary}
\begin{proof}
Write $b_0=\log(\theta/y_0)=-\ell-\log2+1/4+\pi^2/16$.
Then $L_*:=\log(R/a_*)=A\sqrt{k}+b_0-k^{-1/8}$.
The coefficient $b_0$ cancels the order-$\sqrt{k}$ terms in
\eqref{eq:model-spectrum}. Equations \eqref{eq:full-spectrum} and
\eqref{eq:model-spectrum} give
\[
 \pi k\beta(a_*)=\pi k-\frac{8A}{\pi^2}k^{3/8}
                       +O_\theta(k^{1/4}+1)<\pi k
\]
for large $k$, while \eqref{eq:half-spectrum} gives the limit $1/2$.
\end{proof}

\section{The sharp CHF energy asymptotics}\label{sec:energy-last}
\subsection{Evaluation of the deterministic upper bound}
The comparison defect in \eqref{eq:chosen-scale} has compact shell
support; there are no self-consistent exterior tails to estimate.
\begin{lemma}[The explicit comparison source]\label{lem:source}
For $a_*$ in \eqref{eq:chosen-scale},
\begin{align}
 \|d_*\|_{E_*}^2&=\frac{k^3a_*^2}{8\pi^2}(1+o(1)),
 \label{eq:source-energy}\\
 \cX(d_*)&\le C_\theta qk^2a_*^2
       \left(1+\log\frac{k}{a_*}\right)=o(qk^3a_*^2).
 \label{eq:source-exchange}
\end{align}
\end{lemma}
\begin{proof}
On the fixed shell, \eqref{eq:eglob} implies
$e_k(r(t))\ge c_\theta k|t|$. Consequently
\begin{equation}\label{eq:source-envelope}
 0\le d_*(r(t))\le C_\theta\frac{a_*^2}{a_*^2+t^2},\qquad
 \int_{-R}^R d_*(r(t))\,\mathrm{d}t\le C_\theta a_*.
\end{equation}
For fixed $u$, \eqref{eq:xi-HF} and $\log(k/a_*)=O(\sqrt{k})$ give
\[
 \frac{e_k(r(a_*u))}{ka_*}\to |u|,\qquad
 \frac{E_*(r(a_*u))}{ka_*}\to\sqrt{1+u^2},\qquad
 d_*(r(a_*u))\to d_\infty(|u|),
\]
where
\begin{equation}\label{eq:dinfty}
 d_\infty(v)=\frac12\left(1-\frac{v}{\sqrt{1+v^2}}\right),
 \qquad v\ge0.
\end{equation}
The identity
\[
 E_*d_*^2=\frac{k^4a_*^4}{4E_*(E_*+e_k)^2}\quad\text{on }\Omega
\]
and the lower bound for $e_k$ majorize the scaled energy integrand
by $C_\theta(1+|u|)^{-3}$. Extend it by zero off the rescaled shell.
Dominated convergence gives
\[
 \|d_*\|_{E_*}^2=\frac{k^3a_*^2}{2\pi^2}
 \left[2\int_0^\infty\sqrt{1+u^2}\,d_\infty(u)^2
               \,\mathrm{d}u+o(1)\right].
\]
The integral equals $1/8$: with $u=\sinh v$ its integrand times
$\mathrm{d}u$ becomes $e^{-2v}\,\mathrm{d}v/4$.
This proves \eqref{eq:source-energy}.
For exchange apply \eqref{eq:radial-X} and Lemma~\ref{lem:shell-kernel}
to $f(t)=d_*(r(t))$. Its mass is $O(a_*)$ by
\eqref{eq:source-envelope}; the increasing function
$m^2(1+\log(C_\theta k/m))$ gives \eqref{eq:source-exchange}.
The relative bound is $O_\theta(\log(k/a_*)/k)=o(1)$.
\end{proof}

Take the infimum in Proposition~\ref{prop:completion}, use
Corollary~\ref{cor:subcritical}, and drop the nonnegative remainder.
Since $\rho=qk^3/(6\pi^2)$, Lemma~\ref{lem:source} yields
\begin{equation}\label{eq:sharp-upper}
 D(k)\le\frac{q\|d_*\|_{E_*}^2+5\cX(d_*)}{\rho}
        =\frac34a_*^2(1+o(1)).
\end{equation}
Thus the upper limit in Theorem~\ref{thm:main} is at most $C_*$.
This argument did not use existence, radiality, or self-consistency
of a minimizer.

\subsection{One number-preserving trial state on the required scale}
\label{sec:trial-constant}
For the matching lower bound it suffices to evaluate one trial state:
the scale is already selected by the deterministic spectrum. Set
\begin{equation}\label{eq:trials}
 a=a_0(k)=y_0k e^{-A\sqrt{k}},\qquad R=\theta k,\qquad
 L=\log(R/a)=A\sqrt{k}+b_0,\qquad b_0=\log(\theta/y_0).
\end{equation}
The parameter $\theta$ is fixed, and the construction is used for
sufficiently large $k$, so that $L>0$. For $0<v\le R$ define
\[
 g_L(v)=\sin\left(\frac\pi{2L}\min\{\log(R/v),L\}\right),\qquad
 w_a(v)=\frac{a g_L(v)}{2\sqrt{v^2+a^2g_L(v)^2}},
\]
with $w_a(0)=1/2$ and zero continuation for $v>R$.
The actual momentum occupation is
$n_k^{\rm tr}(p)=\nu_a(t(|p|))$, where
\begin{equation}\label{eq:nonlineartrial}
 \nu_a(t)=\frac12\left(1-\operatorname{sgn}(t)
                   \sqrt{1-4w_a(|t|)^2}\right)\quad (|t|\le R),
\end{equation}
and $\nu_a(t)=\one_{\{t<0\}}$ outside. The constant Jacobian
\eqref{eq:volume} and $\nu_a(t)+\nu_a(-t)=1$ on $[-R,R]$
give the particle number exactly.
Write $\delta_a(t)=|\nu_a(t)-\one_{\{t<0\}}|$ and introduce the
reference tail
\[
 f_L(x)=\sin\frac{\pi x}{2L},\qquad
 w_0(v)=\frac a{2v}f_L(\log(R/v))\one_{\{a\le v\le R\}}.
\]

\begin{lemma}[Inner matching]\label{lem:micro}
For the preceding trial profile,
\begin{align}
 \int_{-R}^R(w_a(|t|)-w_0(|t|))\,\mathrm{d}t
 &=a\log2+o(a),\label{eq:wmatch}\\
 \int_{-R}^R|t|[\delta_a(t)-w_0(|t|)^2] \,\mathrm{d}t
 &=a^2\left(\frac{\log2}{2}-\frac14\right)+o(a^2).
 \label{eq:dmatch}
\end{align}
The corresponding absolute integrals, even with the factor
$1+|\log(|t|/a)|$, are bounded by $Ca$ and $Ca^2$, respectively.
Moreover,
\begin{equation}\label{eq:moment2}
 \int_{-R}^R t^2\delta_a(t)
 \left(1+\log\frac{2k}{|t|}\right)\,\mathrm{d}t
 \le C_\theta\left(\frac{a^2R}{L^2}+a^3(1+L)\right).
\end{equation}
\end{lemma}
\begin{proof}
For $a\le v\le R$,
\[
 |w_a(v)-w_0(v)|\le Ca^3/v^3,\qquad
 |\delta_a(v)-w_0(v)^2|\le Ca^4/v^4.
\]
For $v<a$ all profiles are bounded and $w_0=0$.
After $t=au$, these are integrable majorants, including the stated
logarithmic moments. For each fixed $u>0$,
$2w_a(au)\to(1+u^2)^{-1/2}$ and
$\delta_a(au)\to d_\infty(u)$. Hence dominated convergence and
\begin{align}
 \int_0^\infty\left[\frac1{\sqrt{1+u^2}}
                    -\frac{\one_{\{u>1\}}}{u}\right]\mathrm{d}u
 &=\log2,\label{eq:matching1}\\
 2\int_0^\infty u\left[d_\infty(u)
                    -\frac{\one_{\{u>1\}}}{4u^2}\right]\mathrm{d}u
 &=\frac{\log2}{2}-\frac14\label{eq:matching2}
\end{align}
prove the matching formulas. The antiderivatives use
$\operatorname{arsinh}u$ and
$[u^2-u\sqrt{1+u^2}+\operatorname{arsinh}u]/2$.
For \eqref{eq:moment2}, split at $|t|=a$.
On $a\le|t|\le R$, use $\delta_a(t)\le2w_a(|t|)^2$,
$x=\log(R/|t|)$, and $f_L(x)\le Cx/L$ to obtain
\[
\begin{aligned}
 &\int_{a\le|t|\le R}t^2\delta_a(t)
        \left(1+\log\frac{2k}{|t|}\right)\,\mathrm{d}t\\
 &\qquad\le\frac{Ca^2R}{L^2}
      \int_0^L e^{-x}(1+\ell+x)x^2\,\mathrm{d}x
 \le C_\theta\frac{a^2R}{L^2}.
\end{aligned}
\]
On $|t|<a$, use $\delta_a(t)\le1/2$ and
$\log(2k/a)=\ell+L$:
\[
\begin{aligned}
 &\int_{|t|<a}t^2\delta_a(t)
        \left(1+\log\frac{2k}{|t|}\right)\,\mathrm{d}t\\
 &\qquad\le\int_0^a t^2
        \left(1+\log\frac{2k}{t}\right)\,\mathrm{d}t
 =a^3\left(\frac{1+\ell+L}{3}+\frac19\right)
 \le C_\theta a^3(1+L).
\end{aligned}
\]
\end{proof}

\begin{lemma}[Pairing integral]\label{lem:pair}
For the trial profile,
\begin{equation}\label{eq:trial-pairing-integral}
\begin{aligned}
 &\iint_{[-R,R]^2}w_a(|t|)w_a(|s|)
        \log\frac{2k}{|t-s|}\,\mathrm{d}t\,\mathrm{d}s\\
 &\qquad=a^2\left[\frac{2L^3}{\pi^2}
 +\frac{4L^2}{\pi^2}(\ell+2\log2)+o(L^2)\right].
\end{aligned}
\end{equation}
Replacing the logarithm by $H_k(t,s)$ changes this expression by
$o(a^2L^2)$.
\end{lemma}
\begin{proof}
Write $w_a=w_0+\delta w$. On each sign of $t$ set $t=\sigma Re^{-x}$,
$0\le x\le L$, so that $w_0(|t|)|\mathrm{d}t|=(a/2)f_L(x)\,\mathrm{d}x$.
For $s=\tau Re^{-z}$,
\[
 \log\frac{2k}{|t-s|}
 =\min(x,z)+\ell-\log|1-\sigma\tau e^{-|x-z|}|.
\]
The last convolution kernel is in $L^1(\R)$ and contributes
$O(a^2L)$. The two remaining tail--tail terms are
\[
 a^2\left[\frac{2L^3}{\pi^2}+\frac{4\ell L^2}{\pi^2}\right],
\]
because $\int_0^Lf_L=2L/\pi$ and
$\iint f_L(x)\min(x,z)f_L(z)\,\mathrm{d}x\,\mathrm{d}z=2L^3/\pi^2$.
For fixed $u$, the potential of $w_0$ at $t=au$, divided by $aL^2$,
converges to $4/\pi^2$, using $\int_0^Lx f_L(x)\,\mathrm{d}x=4L^2/\pi^2$.
The local logarithmic singularity is integrable uniformly on compact
$u$ intervals. Lemma~\ref{lem:shell-kernel}, applied to $w_0$ whose
mass is $O(aL)$, bounds this divided potential uniformly by a constant.
Thus the majorants of Lemma~\ref{lem:micro} allow integration against
$\delta w(au)$ and removal of the compact restriction. The two cross
terms contribute $8(\log2)a^2L^2/\pi^2+o(a^2L^2)$.
Since $|\delta w(au)|\le C(1+u)^{-3}$ for $u\ge0$, the
correction--correction term is $O(a^2L)$.

For the exact geometry use \eqref{eq:shell-geometry}.
In the tail--tail part a factor $|t|/k=\theta e^{-x}$ yields
\[
 C_\theta a^2\int_0^L e^{-x}f_L(x)
        \int_0^L f_L(z)(\min(x,z)+1)\,\mathrm{d}z\,\mathrm{d}x
 =O_\theta(a^2),
\]
by $f_L(x)\le Cx/L$ and
$\int f_L(z)\min(x,z)\,\mathrm{d}z\le CxL$.
The convolution remainder and terms involving $\delta w$ are
$O_\theta(a^2L)$, using its absolute and logarithmic moments from
Lemma~\ref{lem:micro}. All are $o(a^2L^2)$.
\end{proof}

\begin{proposition}[Matching trial bound]\label{prop:sharp-trial}
The admissible occupation $n_k^{\rm tr}$ satisfies
\begin{equation}\label{eq:trial-energy-final}
 \cE(n_k^{\rm tr})-\cE(P_k)
 =-\frac{qk^3a_0(k)^2}{8\pi^2}(1+o(1)).
\end{equation}
\end{proposition}
\begin{proof}
Put $h^{\rm tr}=n_k^{\rm tr}-P_k$.
Since the trial occupation has the prescribed particle number,
the exact decomposition \eqref{eq:energy-start} gives
\[
 \cE(n_k^{\rm tr})-\cE(P_k)
 =q\int e_k|h^{\rm tr}|\,\dbar^3p
  -\cX(h^{\rm tr})-\cX(w_{n_k^{\rm tr}}).
\]
Thus the kinetic contribution and the linear ordinary-exchange
contribution are evaluated together.
For $0<|t|\le R=\theta k$, the explicit dispersion
\eqref{eq:xi-HF} yields
\[
\begin{aligned}
 e_k(r(t))
 &=k|t|+\frac{|t|}{\pi}\log\frac{2k}{|t|}\\
 &\quad+O_\theta\left(
 t^2\left[1+\frac1k\log\frac{2k}{|t|}\right]\right).
\end{aligned}
\]
Lemma~\ref{lem:micro} and \eqref{eq:sinmoment} imply
\[
\begin{aligned}
 \int_{-R}^R|t|\delta_a(t)\,\mathrm{d}t
 &=\frac{a^2}{4}\bigl[L+2\log2-1+o(1)\bigr],\\
 \int_{-R}^R|t|\delta_a(t)
       \log\frac{2k}{|t|}\,\mathrm{d}t
 &=\frac{a^2L^2}{2}
      \left(\frac14+\frac1{\pi^2}\right)+O_\theta(a^2L).
\end{aligned}
\]
For the second identity, replace $\delta_a$ by $w_0^2$ and use
$\log(2k/|t|)=\ell+x$, where $x=\log(R/|t|)$.
The replacement error is $O_\theta(a^2L)$ by the absolute logarithmic
moment bound in Lemma~\ref{lem:micro}.
By \eqref{eq:moment2}, the integrated dispersion remainder is bounded by
\[
 C_\theta\left(\frac{a^2R}{L^2}+a^3(1+L)\right)=o(ka^2).
\]
Using the exact volume Jacobian therefore gives
\[
\begin{aligned}
 q\int e_k|h^{\rm tr}|\,\dbar^3p
 &=\frac{qk^2}{2\pi^2}\int_{-R}^R
      e_k(r(t))\delta_a(t)\,\mathrm{d}t\\
 &=\frac{qk^3a^2}{8\pi^2}
    \left[L+2\log2-1
    +\frac{2L^2}{\pi k}\left(\frac14+\frac1{\pi^2}\right)
    +o(1)\right].
\end{aligned}
\]
The volume-coordinate absolute mass of $h^{\rm tr}$ is $O(a)$.
Thus Lemma~\ref{lem:shell-kernel} gives
\[
 0\le\cX(h^{\rm tr})
 \le C_\theta qk^2a^2\left(1+\log\frac{k}{a}\right)
 =o(qk^3a^2).
\]
For the pairing term, \eqref{eq:radial-X} and Lemma~\ref{lem:pair} give
\[
 \cX(w_{n_k^{\rm tr}})
 =\frac{qk^2a^2}{4\pi^3}
 \left[\frac{2L^3}{\pi^2}
        +\frac{4L^2}{\pi^2}(\ell+2\log2)+o(L^2)\right].
\]
Combining the three terms yields
\begin{equation}\label{eq:trial-energy-accounting}
\begin{aligned}
 &\frac{8\pi^2}{qk^3a^2}
        [\cE(n_k^{\rm tr})-\cE(P_k)]\\
 &\quad=L-\frac{4L^3}{\pi^3k}+2\log2-1
 -\frac{8L^2}{\pi^3k}(\ell+2\log2)
 +\frac{2L^2}{\pi k}\left(\frac14+\frac1{\pi^2}\right)+o(1).
\end{aligned}
\end{equation}
With $L=A\sqrt{k}+b_0$ and
$b_0=-\ell-\log2+1/4+\pi^2/16$, the right side tends to $-1$.
This proves \eqref{eq:trial-energy-final}.
\end{proof}

\begin{proof}[Proof of Theorem~\ref{thm:main}]
The variational principle and Proposition~\ref{prop:sharp-trial} give
$D(k)\ge\tfrac34a_0(k)^2(1+o(1))$.
Equation \eqref{eq:sharp-upper} gives the converse inequality with
$a_*$ in place of $a_0$. Since $a_*/a_0\to1$, the two bounds agree,
and $3y_0^2/4=C_*$ gives \eqref{eq:constant-target}.
In particular the exponential rate and polynomial power follow
without a preliminary two-sided-scale proposition.
\end{proof}

\section{The occupation-defect profile}\label{sec:profile}
\begin{corollary}[Weighted occupation-defect profile]\label{cor:L2-profile}
Let $n_k$ be the minimizer and write
$\widetilde h_k(t)=n_k(r(t))-\one_{\{t<0\}}$.
Then
\begin{equation}\label{eq:global-weighted}
 \int_{\R^3}E_*(p)
 \bigl(|n_k(p)-P_k(p)|-d_*(p)\bigr)^2\,\dbar^3p
 =o(k^3a_*^2).
\end{equation}
Equivalently, with
$\widehat E_k(u)=E_*(r(a_*u))/(ka_*)$, the same convergence reads
\[
 \int_{-k/(3a_*)}^{\infty}\widehat E_k(u)
 \bigl(|\widetilde h_k(a_*u)|-d_*(r(a_*u))\bigr)^2
 \,\mathrm{d}u\longrightarrow0.
\]
For a fixed-weight version, define the shell-restricted profile on $\R$
by
\[
 F_k(u)=
 \begin{cases}
 |\widetilde h_k(a_*u)|,&|u|\le R/a_*,\\
 0,&|u|>R/a_*.
 \end{cases}
\]
Then
\begin{equation}\label{eq:fixed-weight}
 \boxed{\int_\R\sqrt{1+u^2}
 |F_k(u)-d_\infty(|u|)|^2\,\mathrm{d}u\longrightarrow0.}
\end{equation}
The energy-space convergence \eqref{eq:global-weighted} also holds
for arbitrary (possibly nonradial) $n_k\in\mathcal A_\rho$ satisfying
$\cE(n_k)-\inf_{\mathcal A_\rho}\cE=o(qk^3a_*^2)$.
The radial fixed-weight conclusion holds for such approximate
minimizers whenever they are radial.
\end{corollary}
\begin{proof}
For the minimizer, Proposition~\ref{prop:completion} yields
\[
 \frac q4\bigl\||n_k-P_k|-d_*\bigr\|_{E_*}^2
 \le q\|d_*\|_{E_*}^2+5\cX(d_*)-\rho D(k).
\]
Lemma~\ref{lem:source} and Theorem~\ref{thm:main} show that the
right-hand side is $o(qk^3a_*^2)$. This proves the global assertion
and, after the exact volume-coordinate transformation, its rescaled
version. On $|t|\le R$, \eqref{eq:eglob} implies
\[
 \frac{E_*(r(a_*u))}{ka_*}\ge c_\theta\sqrt{1+u^2}.
\]
The deterministic profiles $d_*(r(a_*u))$, extended by zero outside
$|u|\le R/a_*$, converge to $d_\infty(|u|)$ in the fixed weighted
space by \eqref{eq:source-envelope} and dominated convergence.
The triangle inequality proves \eqref{eq:fixed-weight}.
The proof for approximate minimizers is identical.
\end{proof}

\begin{remark}[Why the global weight is specified]
Without a shell cutoff, one should not replace $\widehat E_k(u)$
by $\sqrt{1+u^2}$ over the entire positive half-line without another
argument. Far beyond the Fermi shell, $t(r)\asymp r^3/k^2$, whereas
$E_*(r)\asymp r^2$. Thus these weights are not globally comparable
in the direction needed for that stronger statement. The global
energy-weighted assertion and the fixed-weight shell assertion above
follow directly from completion of the square; no self-consistent
exterior-tail analysis is needed for either.
\end{remark}

\section{Comparison with quantum-mechanical predictions}
\label{sec:quantum-comparison}
Theorems~\ref{thm:mueller-main} and \ref{thm:main} concern the
specified density-matrix functionals. Their physical interpretation
requires a separate comparison with the interacting quantum gas. We
make that comparison at the level of energy per particle and explicitly
distinguish rigorous functional asymptotics, rigorous quantum results in
scaled regimes, and the thermodynamic high-density prediction.

\subsection{A common normalization and the quantum benchmark}
For the physical unpolarized gas take $q=2$ and introduce the Seitz radius
(in Bohr radii)
\begin{equation}\label{eq:rs-normalization}
 r_s=\left(\frac{3}{4\pi\rho}\right)^{1/3},\qquad
 k=\frac{\kappa_0}{r_s},\qquad
 \kappa_0=\left(\frac{9\pi}{4}\right)^{1/3}.
\end{equation}
Let $e_{\rm Q}(r_s)$ denote the thermodynamic ground-state energy per
particle of neutral Coulomb jellium in the unpolarized sector. The
standard high-density quantum prediction is
\begin{equation}\label{eq:GMB-benchmark}
 e_{\rm Q}(r_s)-e_{\rm HF}^{\rm hom}(r_s)
 =a_{\rm Q}\log r_s+B_{\rm Q}+o(1),\qquad
 a_{\rm Q}=\frac{1-\log2}{\pi^2}.
\end{equation}
This is the Macke--Gell-Mann--Brueckner correlation formula; its
leading logarithm is obtained by the resummation of the long-range
screening contributions~\cite{Macke,Pines,GMB}. The Hartree convention
and the Hartree--Fock reference are also discussed in~\cite{LG}.
The coefficient is $a_{\rm Q}=0.0310906908\ldots$ in Hartree units;
in Rydberg units it is twice as large. The constant $B_{\rm Q}$ is not
needed below. Since $\log r_s=\log\kappa_0-\log k$, the predicted
positive energy gain is
\[
 D_{\rm Q}(k):=e_{\rm HF}^{\rm hom}(k)-e_{\rm Q}(k)
 =a_{\rm Q}\log k+O(1).
\]
All statements involving $D_{\rm Q}$ at this precision are understood
relative to the physical benchmark \eqref{eq:GMB-benchmark}, not as
additional theorems proved in this paper.

For clarity, the three results in the same units are summarized in
Table~\ref{tab:comparison}. No factor of density is included in any
entry. A per-volume correlation is obtained by multiplication by $\rho$.
\begin{table}[htbp]
\centering
\small
\renewcommand{\arraystretch}{1.35}
\begin{tabularx}{\textwidth}{@{}>{\raggedright\arraybackslash}p{0.18\textwidth}>{\raggedright\arraybackslash}X >{\raggedright\arraybackslash}p{0.27\textwidth}@{}}
\toprule
Model & Gain per particle over the filled Fermi sea & Status here\\
\midrule
M\"uller & $\displaystyle \frac{3}{4\pi^2}\log k+O(1)$
 & Theorem~\ref{thm:mueller-main}\\
CHF & $\displaystyle C_* k^2e^{-\pi^{3/2}\sqrt{k}}(1+o(1))$
 & Theorem~\ref{thm:main}\\
Coulomb quantum gas ($q=2$)
 & $\displaystyle \frac{1-\log2}{\pi^2}\log k+O(1)$
 & Thermodynamic high-density prediction \eqref{eq:GMB-benchmark}\\
\bottomrule
\end{tabularx}
\caption{Correlation scales in Hartree units, with
$C_*=12e^{-1/2-\pi^2/8}$. The status of the quantum entry differs
from that of the two functional theorems.}
\label{tab:comparison}
\end{table}

In the traditional density variable, our functional results become
\begin{align}
 e_{\rm M}^{\rm hom}(r_s)-e_{\rm HF}^{\rm hom}(r_s)
 &=\frac{3}{4\pi^2}\log r_s+O(1),\label{eq:M-rs}\\
 e_{\rm CHF}^{\rm hom}(r_s)-e_{\rm HF}^{\rm hom}(r_s)
 &=-C_*\kappa_0^2r_s^{-2}
 \exp\left(-\frac{\pi^{3/2}\sqrt{\kappa_0}}{\sqrt{r_s}}\right)
 (1+o(1)).\label{eq:CHF-rs}
\end{align}
Thus the M\"uller coefficient is
\[
 a_{\rm M}=\frac{3}{4\pi^2}=0.0759908877\ldots,
 \qquad
 \frac{a_{\rm M}}{a_{\rm Q}}
 =\frac{3}{4(1-\log2)}=2.4441685149\ldots.
\]
Relative to the quantum prediction, M\"uller therefore overestimates
the leading correlation magnitude by this factor. CHF, in contrast,
misses the logarithmic scale: conditionally on
\eqref{eq:GMB-benchmark},
\begin{equation}\label{eq:quantum-ratios}
 \frac{D_{\rm M}(k)}{D_{\rm Q}(k)}
 \longrightarrow\frac{3}{4(1-\log2)},\qquad
 \frac{D(k)}{D_{\rm Q}(k)}\longrightarrow0.
\end{equation}
There is no inconsistency with their common leading total energy:
both corrections are $o(k)$, and hence leave the kinetic $k^2$ and
exchange $k$ terms unchanged. Agreement at those orders is therefore
a weak test of the correlation mechanism.

Cs\'anyi and Arias~\cite{CA} reported a fitted logarithmic coefficient
$0.0570$ for CH/M\"uller in a numerical study covering $0.01<r_s<30$.
This range describes the densities studied, not a specified fitting
interval. Such a finite-range fit does not determine the limiting
coefficient $3/(4\pi^2)=0.07599\ldots$. Our theorem gives that
coefficient with a bounded remainder, not a limit for the remainder;
it does not attribute the discrepancy to particular finite-range or
numerical errors.

To illustrate the CHF scale, denote the positive leading term in
\eqref{eq:CHF-rs} by
\[
 D_{\rm lead}(r_s)
 =C_*\kappa_0^2r_s^{-2}
  \exp\left(-\frac{\pi^{3/2}\sqrt{\kappa_0}}{\sqrt{r_s}}\right).
\]
Evaluating this expression gives
$D_{\rm lead}(0.04)\approx8.7\times10^{-14}$ Hartree,
$D_{\rm lead}(0.04)/e_{\rm HF}^{\rm hom}(0.04)
 \approx1.3\times10^{-16}$, and
$D_{\rm lead}(0.01)\approx2.5\times10^{-29}$ Hartree.
These are leading-term evaluations, not certified finite-density minima.
The first relative value is comparable to double-precision roundoff,
so subtracting total energies is poorly conditioned. Higher precision
or a direct calculation of the difference can improve numerical
resolution; neither replaces analytic control of the asymptotic law.

\subsection{Evidence for the M\"uller lower-bound conjecture}
\label{sec:mueller-lower-bound}
The comparison has a second interpretation, distinct from judging the
accuracy of an approximation. A motivating question is whether the
M\"uller ground-state energy is a lower bound to the exact quantum
energy for the same Coulomb system~\cite{FLSS,Siedentop}. For this
question, a larger correlation magnitude has the favorable sign.
For the unpolarized homogeneous gas ($q=2$), Theorem~\ref{thm:mueller-main}
and the quantum benchmark \eqref{eq:GMB-benchmark} give
\begin{equation}\label{eq:M-quantum-separation}
 \boxed{
 e_{\rm Q}(k)-e_{\rm M}^{\rm hom}(k)
 =\frac{\log2-\tfrac14}{\pi^2}\log k+O(1).
 }
\end{equation}
This statement is conditional on \eqref{eq:GMB-benchmark}.

The spin multiplicity is essential in this comparison. For fixed $q$
equally populated components, counting the components in the RPA ring
expression~\cite{GMB} gives the logarithmic coefficient per particle
\[
 a_{\rm RPA}(q)=\frac{q(1-\log2)}{2\pi^2},
\]
whereas $a_{\rm M}=3/(4\pi^2)$ is independent of $q$.
Hence, under the corresponding $q$-component quantum high-density
benchmark,
\[
 e_{\rm Q}^{(q)}(k)-e_{\rm M}^{{\rm hom},(q)}(k)
 =\frac{3-2q(1-\log2)}{4\pi^2}\log k+O_q(1).
\]
The leading coefficient is positive for $q\le4$ and negative for
$q\ge5$. Thus this benchmark supports the proposed lower-bound ordering
in the physical case $q=2$, but predicts the opposite ordering for
sufficiently large $k$ when $q\ge5$. An extension of the lower-bound
conjecture to arbitrary spin multiplicities would therefore be
incompatible with this high-density benchmark.

By Proposition~\ref{prop:M-periodic}, this comparison is not an artifact
of restricting M\"uller to homogeneous spin-symmetric states:
convexity and averaging leave the periodic minimum unchanged, and
$e_{\rm M}^{{\rm hom},(q)}(k)$ equals the unrestricted periodic
thermodynamic energy per particle. Under the same quantum
high-density benchmark, allowing inhomogeneous or
spin-polarized density matrices cannot restore a high-density lower
bound for $q\ge5$, whereas for $q=2$ the positive logarithmic separation
\eqref{eq:M-quantum-separation} compares the quantum benchmark with the
unrestricted periodic M\"uller thermodynamic energy.

Returning to $q=2$, since $\log2-1/4>0$, the predicted separation is
strictly positive for all sufficiently large $k$ and is itself of
logarithmic order; a bounded remainder cannot reverse its sign in that
limit. Thus the analytic M\"uller coefficient supplies additional
quantitative evidence for the lower-bound conjecture, not merely a
measure of overcorrelation. It does not establish the conjecture for
arbitrary densities, particle numbers, atoms, or molecules, and it
does not strengthen the mathematical status of the quantum benchmark.
No such quantum statement is used in either functional proof.

\subsection{What is rigorously known for the quantum gas}
The exchange-order control in the high-density quantum problem belongs
to the correlation-estimate framework of Graf and Solovej~\cite{GS}.
Resolving the much smaller logarithmic correlation term requires
additional arguments. For weak, sufficiently regular interactions in
mean-field scaling, Benedikter, Nam, Porta, Schlein, and
Seiringer~\cite{BNPSS} obtained the random-phase correlation energy.
Christiansen, Hainzl, and Nam extended the rigorous upper and matching
lower analysis to singular interactions including Coulomb
\cite{CHNupper,CHNlower}. Their result establishes the corresponding
Gell-Mann--Brueckner analogue, including the next order, in a scaled
many-fermion problem. It does not by itself establish
\eqref{eq:GMB-benchmark} for unscaled thermodynamic jellium.
The latter distinction, rather than the formal similarity of the
logarithms, determines what can be used as a proved quantum comparison.

There is a second distinction concerning the reference state.
Gontier, Hainzl, and Lewin~\cite{GHL} proved that the high-density
Hartree--Fock energy improvement from allowing translation-symmetry
breaking is exponentially small. Thus it cannot change the leading
quantum logarithm. Our reference in both functional theorems is,
however, exactly the uniform spin-symmetric Fermi sea. At the
exponentially small CHF scale, their result cannot simply be discarded
or identified with our exponent and prefactor. We neither identify the
unrestricted Hartree--Fock minimizer with $P_k$ nor analyze CHF in that
larger variational class. By contrast, Proposition~\ref{prop:M-periodic}
shows that restricting periodic M\"uller jellium to translation--spin
invariant states does not change its minimum at any density; an
invariant global minimizer can be chosen.

\subsection{Why the two approximate mechanisms differ}
The distinction is visible directly in the variational structure. For
M\"uller, the pinned core has $b=\sqrt n=1$. Its exchange field is
nonzero at the Fermi surface, with
$\sigma_{P_k}(k)=k/\pi$. The fractional tail is therefore driven by
an already present core source. In the region
$1\ll r-k\ll k$, the equation $2gb=\sigma_b$ produces an inverse
energy tail. Its integrated contribution is the logarithm in
\eqref{eq:M-Q}. The energy cancellation in
\eqref{eq:M-energy-identity} fixes its coefficient; counting only the
exchange defect would incorrectly double it.

For CHF, $w_n=\sqrt{n(1-n)}$ vanishes on a filled Fermi sea,
so there is no analogous nonzero core source. The deterministic
comparison identifies the scale at which a fractional occupation
can lower the energy. The angular Coulomb logarithm and the normal
Fermi-surface coordinate produce the Green kernel $\min(x,y)$,
whose first eigenvalue is $4L^2/\pi^2$. The full comparison norm
reaches order one at $L\sim\pi^{3/2}\sqrt{k}/2$, whereas each
one-sided norm is then close to one half. The latter fact absorbs
ordinary exchange, and the completed square makes the energy gain
quadratic in the shell width. The linear spectral correction and
the nonlinear trial core determine $C_*$. This explanation uses no
asymptotic formula for the actual self-consistent gap.

These are mechanisms of the approximate functionals, not a derivation
of the quantum screening diagrams from them. In particular the
auxiliary CHF quantity $\Delta=Vw_n$ is not proved to be a many-body
excitation gap, and the Hartree--Fock--Bogoliubov representation
\cite{SS} is not evidence for superconductivity in the present quantum
comparison.

\subsection{Scope, atomic results, and functional design}
The pointwise ordering
$\cE_{\rm M}\le\cE_{\rm CHF}\le\cE_{\rm HF}$ in
\eqref{eq:functional-order} orders these approximations, not the
exact quantum energy. One-particle representability does not make an
approximate two-particle interaction functional variational for the
many-body Hamiltonian. Neither an unconditional inequality
$e_{\rm M}\le e_{\rm Q}$ for all densities nor
$e_{\rm Q}\le e_{\rm CHF}$ is proved here. The overcorrelation and
undercorrelation conclusions, as well as the lower-bound evidence in
Section~\ref{sec:mueller-lower-bound}, are leading-order comparisons to
\eqref{eq:GMB-benchmark} in the specified unpolarized sector.

Likewise, the atomic agreement through order $Z^{5/3}$ in
\cite{Siedentop2009,Siedentop} does not determine terms on a smaller
correlation scale such as $Z\log Z$. The homogeneous formulas cannot
be inserted into an atomic local-density expression and then declared
an atomic theorem: uniform semiclassical and localization errors would
have to be controlled separately. No such transfer is used or proved
in this paper.

The results separate three tests: leading total energy, correlation
scale, and correlation coefficient. They also distinguish approximation
accuracy from a favorable sign for a lower bound. A finite-density fit,
a contraction constraint, or an auxiliary gap representation alone
does not ensure the correct high-density correlation. Both functional
limits concern nonrelativistic models. For CHF, translation invariance
and spin symmetry remain part of the variational restriction. For
M\"uller, the same energy is the unrestricted periodic thermodynamic
energy by Proposition~\ref{prop:M-periodic}; the thermodynamic limit is
taken at fixed density before the high-density limit.

\begingroup
\raggedright

\endgroup

\end{document}